\documentclass[fleqn,usenatbib,useAMS]{mnras}

\usepackage{graphicx}	
\usepackage{epstopdf}
\usepackage{subfigure}
\usepackage{amsmath}	
\usepackage{amssymb}	
\usepackage{multicol}        
\usepackage{bm}		
\usepackage{pdflscape}	

\usepackage{colortbl}  
\usepackage{xcolor}
\usepackage{array}
\usepackage{verbatim}

\newcommand{\ha}{H${\alpha}$}
\newcommand{\hb}{H${\beta}$}

\usepackage[T1]{fontenc}
\usepackage{ae,aecompl}

\usepackage{threeparttable} 
\usepackage{makecell}

\title{Testing Colour-magnitude Pattern as A Method in the Search for Changing-Look AGNs}

\author[Zhu et al.]{\Large\textnormal{
	Li-Tao Zhu$^{1}$,
	Zhongxiang Wang$^{1,2}$\thanks{E-mail: wangzx20@ynu.edu.cn},
	P. U. Devanand$^{3,4}$,
	Alok C. Gupta$^{3}$,
	Karan Dogra$^{3,4}$,
	Jie Li$^{5,6}$}\\ \\ \Large\textnormal{
	Ju-Jia Zhang$^{7,8}$,
	Shun-Hao Ji$^{1}$,
	Si-Si Sun$^{1}$	}
	\\
	$^{1}$Department of Astronomy, School of Physics and Astronomy, Yunnan University, Kunming 650091, China; wangzx20@ynu.edu.cn\\
	$^{2}$Shanghai Astronomical Observatory, Chinese Academy of Sciences, Shanghai 200030, China\\
	$^{3}$Aryabhatta Research Institute of Observational Sciences (ARIES), Manora Peak, Nainital-263001, India\\
	$^{4}$Department of Applied Physics/Physics, Mahatma Jyotiba Phule Rohilkhand University, Bareilly-243006, India\\
	$^{5}$Key Laboratory for Research in Galaxies and Cosmology, Department of Astronomy, University of Science and Technology of China,\\
	Hefei 230036, China\\
	$^{6}$School of Astronomy and Space Science, University of Science and Technology of China, Hefei 230026, China\\
	$^{7}$Yunnan Observatories, Chinese Academy of Sciences, Kunming 650216, China\\
	$^{8}$Key Laboratory for the Structure and Evolution of Celestial Objects, Chinese Academy of Sciences, Kunming 650216, China
}

\date{Accepted XXX. Received YYY; in original form ZZZ}

\pubyear{2024}

\begin{document}
	\label{firstpage}
	\pagerange{\pageref{firstpage}--\pageref{lastpage}}
	
	\maketitle
	
	\begin{abstract}

We develop a simple method to search for changing-look (CL) active 
galactic nucleus (AGN) candidates, and conduct a test run.  
In this method, optical variations of AGNs are monitored and CL-AGNs 
may appear to have a pattern of being bluer when in brightening flare-like 
events. Applying this method, previously-classified type 2 AGNs 
that show the bluer-when-brighter (BWB) pattern are selected. Among more than 
ten thousands type 2 AGNs classified
in the Sloan Digital Sky Survey (SDSS), we find 73 candidates with possibly
	the strongest BWB pattern. We note that 13 of them have previously 
	been reported as CL-AGNs. We have observed nine candidates, and 
found that five among them showed the CL transition from type 2 to type 1. 
In addition,
we also test extending the selection to previously-classified type 1 AGNs in
the SDSS by finding sources with a possible redder-when-brighter pattern, 
but none of the three sources observed by us is found to show the 
transition from type 1 to type 2. We discuss the variation properties 
in both the success and failure cases, and plan to observe more 
	candidates selected with the method.
	From the observational results, a detailed comparison between 
		the CL-AGNs and none CL-AGNs will  
		help quantitatively refine the selection criteria and in turn
		allow us to configure the general properties of CLAGNs.
	\end{abstract}
	
	\begin{keywords}
		galaxies: active --- quasars: emission lines
	\end{keywords}
	
	
	
	
	\section{INTRODUCTION}

Active Galactic Nuclei (AGNs) are among the most luminous and dynamic objects
in the universe, powered by accretion onto super-massive black holes (SMBHs) at
the centers of galaxies. A unification scheme for understanding AGNs'
appearances, structural components, and physical properties has been established
since the late 1980s (e.g., \citealt{law87, ant93, up95, tad08}). In
the unified model, there are mainly two types of AGNs, type 1 and type 2, whose
classifications depend on whether or not we can view the central engine and 
the broad-line region (BLR) and consequently on whether or not broad emission 
lines (BELs) are observed. Type 1 AGNs typically exhibit prominent broad
($\gtrsim$ 1000\,km\,s$^{-1}$) emission lines in their optical and
ultraviolet (UV) 
emission, while type 2 AGNs only show narrow lines ($<$ 1000\,km\,s$^{-1}$). 
There are also types 1.2, 1.5, 1.8, and 1.9 \citep{Win92}, classified based on 
the ratio of the strength of H$\beta$ to that of [O III] 5007; the weaker 
the H$\beta$ line, the higher the type number. The unified model has stood 
as the framework 
for AGNs, withstanding numerous tests and supported by a multitude of 
observational evidence. However, recent observations of the Changing-Look 
phenomenon in AGNs show challenges or at least raise questions about the unified
model. 

The so-called changing-look (CL) may be simply
characterized by appearances or disappearances of BELs in 
optical/UV spectra taken at different epochs 
(e.g., \citealt{to76, crp+86, sbw93, ajk+99, eh01, ddc+14, spg+14, lcm+15}).
The drastic changes in BELs would indicate AGN type transitions between
type 1 and type 2 or other intermediate types.
The thus-found CL AGNs (CLAGNs) may also be referred to as 
changing-state AGNs (CSAGNs; \citealt{grs+20,rt22}) in order to
distinguish them from those found in X-rays, the 
changing-obscuration AGNs (COAGNs; for details about the latter, see, e.g., 
\citealt{mbc+21,rt22}).
In the optical/UV, discoveries of many CLAGNs are enabled by
the Sloan Digital Sky Survey (SDSS) and typical timescales of the 
CL activities are $< 10$--20\,yr, limited by the taking times of
at least two spectra for comparison. Extreme cases of monthly CL
timescales have also been found (e.g., \citealt{tra+19, kcb+19, zte+22}).
How to explain such short changing timescales in the unified model becomes
a problem to be discussed, as intrinsic changes in the accretion of AGNs
should follow the viscous timescale of the accretion disc, which would be of 
the order of $\sim 10^4$\,yr (see discussion in, e.g., 
\citealt{mrl16, rfg18, nd18, rae+19}). 
Different possible scenarios have been proposed, such as changes in
the innermost regions of the accretion disc on the thermal and
heating/cooling front timescales \citep{smg18}, magnetically supported 
thick disc \citep{db19}, magnetic accretion disc with outflows 
\citep{fcl+21}, or radiation pressure instability occuring in the 
narrow ring between the outer standard disc and the inner advection-dominated 
accretion flow (ADAF; \citealt{scb20}).

To understand the physical processes that produce the CL phenomenon, properties
and distinctions of CLAGNs should be thoroughly probed. Efforts have been 
made to find
more CLAGNs and build a large sample for property studies
(e.g., \citealt{rac+16, rcr+16, ghc+17, ywf+18, rgc+20, zlw+24}). 
There are roughly two types of methods used for systematically 
searching for CLAGNs, spectrum-based and light-curve--based. 
The first involves comparing spectra obtained from either the same survey 
(e.g., \citealt{gpa+22, zte+24}) or from different surveys 
(e.g., \citealt{ywf+18, dzg+24, gzf+24, gzg+24}).
This type can quickly and efficiently identify CLAGNs, but may also miss 
many of them because the results are highly dependent on the plans of the
spectroscopy surveys and these surveys are time consuming.
The second one tries to draw characteristics of AGN variabilities,
identify the candidate CLAGNs through certain selection criteria in
optical (e.g., \citealt{mrl+16, fgg+19, grs+20, lmb+22, lsp+23, wwg+24}) 
or infrared 
(e.g., \citealt{swj+20, wzb+23}), in some of which the machine learning 
techniques are applied, and follow with spectroscopic confirmation. 
This type is enabled by the availability of rich amounts of light-curve data 
at multi-bands from different photometric surveys.

In our initial study of
AGN variation patterns, we found four CLAGNs \citep{zlw+24}, and realized
that they likely shared a similar pattern of being bluer when in brightening
flare-like events. This type of bluer-when-brighter (BWB) behaviour has been 
known in AGNs and was noticed in CLAGNs (see \citealt{ywf+18} and references therein).
These variations are often seen in type 1 AGNs, but not in type 2s, 
since the latter typically have weak variations. Sources showing a BWB 
pattern among type 2 AGNs might have undergone the CL transitions.
We thus conducted a follow-up study to explore a method of finding 
CLAGNs with the BWB pattern. We essentially went through type 2 AGNs identified
in the SDSS database and selected those with stronger BWB variations as the
targets for finding CLAGNs. It should be pointed out that our method
relies on the recent light-curve data provided by large photometric surveys,
and the CL transitions could occur recently, thus reflected by the data.
In this reported work, we obtained spectra for nine targets, and five
of them have been identified as
CLAGNs. Among the identified, three are newly discovered and two were 
reported in \citet{wwg+24}. The latter two are 
J1020+2437 and J1150+3503 (see Table~\ref{tab:info1});
our spectroscopic 
observations were conducted before the appearance of \citet{wwg+24} and
we were not aware of the identification at the time. In addition, we also
selected a few type 1 AGNs to test if their non-BWB behaviour would indicate
the transition from type 1 to type 2. We observed three of the selected sources.

We report our observational results in this paper. In Section~\ref{sec:tgt},
we describe the target selection method. In Section~\ref{sec:obs}, 
we provide information for our spectroscopic observations with the 2.4-m 
LiJiang Telescope (LJT, \citealt{wan+19}) and the
3.6-m Devasthal Optical Telescope (DOT, \citealt{dot18}) and 
the related data reduction process. We present the analysis of the photometric
data and spectra for nine type-2 and three type-1 targets and 
the identification of 5 CLAGNs 
in Section~\ref{sec:ar}. The results are discussed and summarized 
in Section~\ref{sec:ds}.  Throughout this paper, we adopted cosmological 
parameters from the Planck mission \citep{paa+18}, 
with \emph{$H_0$} = 67\,km\,s$^{-1}$\,Mpc$^{-1}$ and $\Omega_m$ = 0.32.

	
\section{Target selection method}	
\label{sec:tgt}
	\begin{figure}
		\centering
		\includegraphics[width=0.79\linewidth]{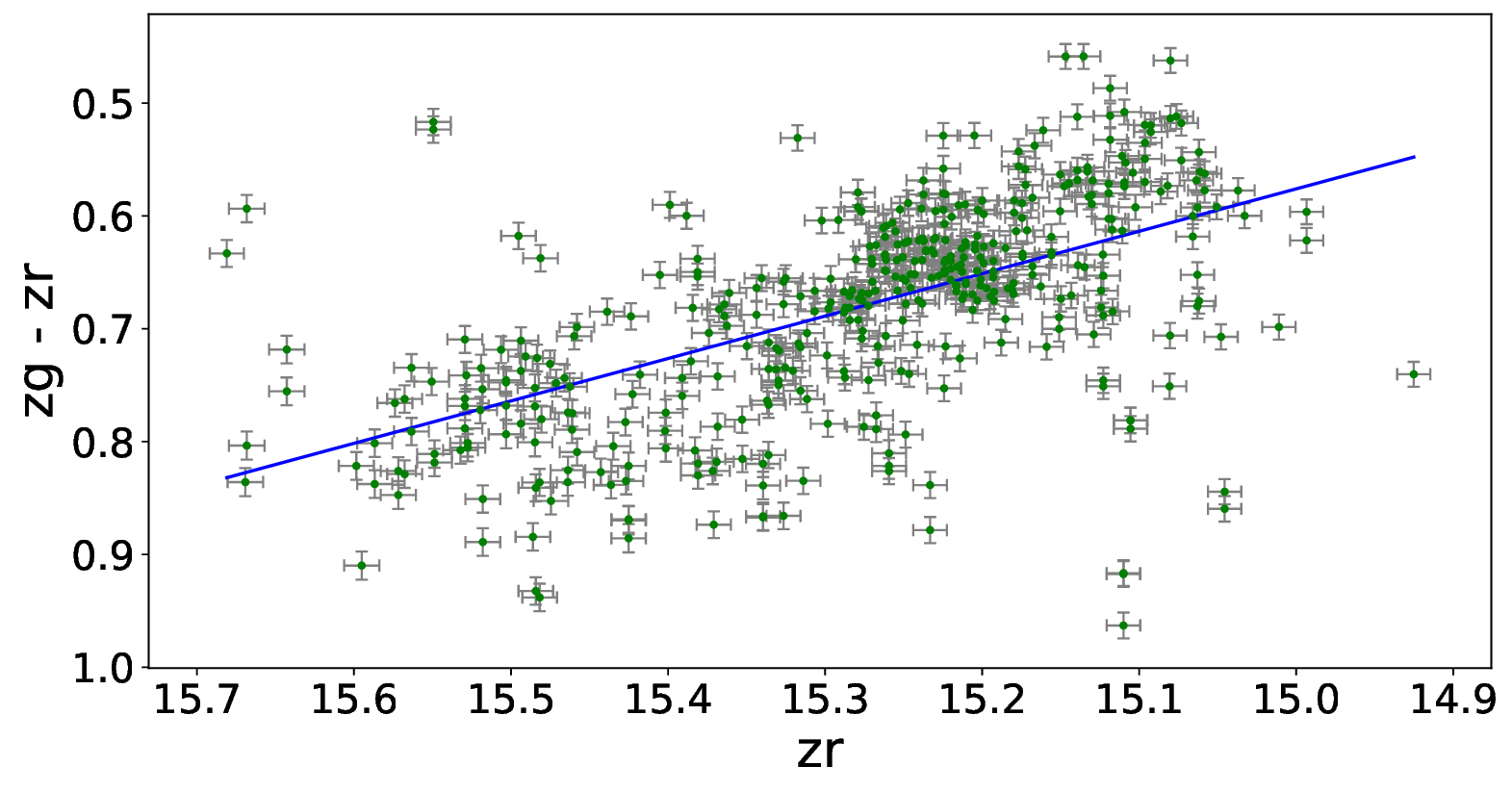}
		\includegraphics[width=0.79\linewidth]{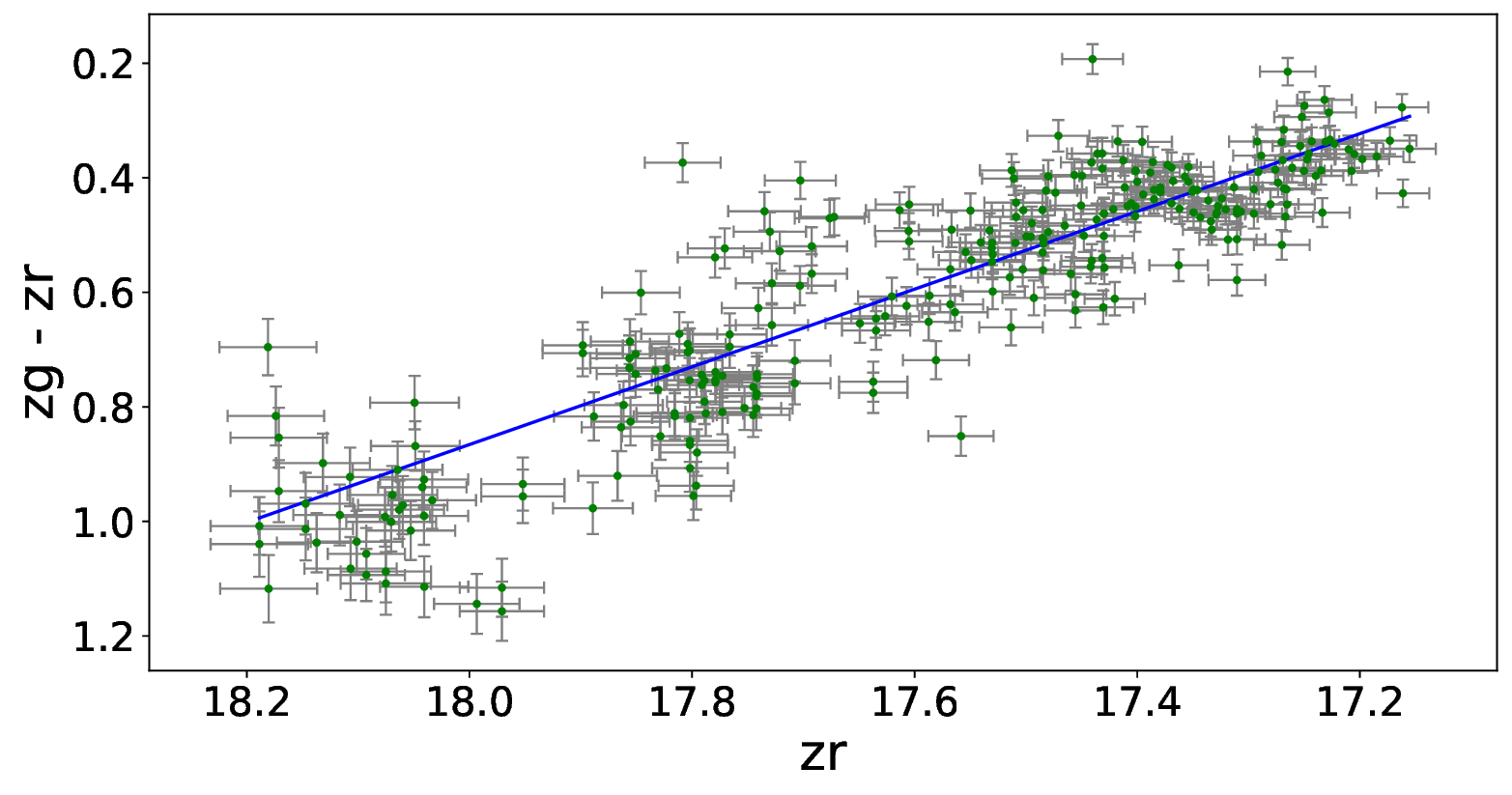}
		\includegraphics[width=0.79\linewidth]{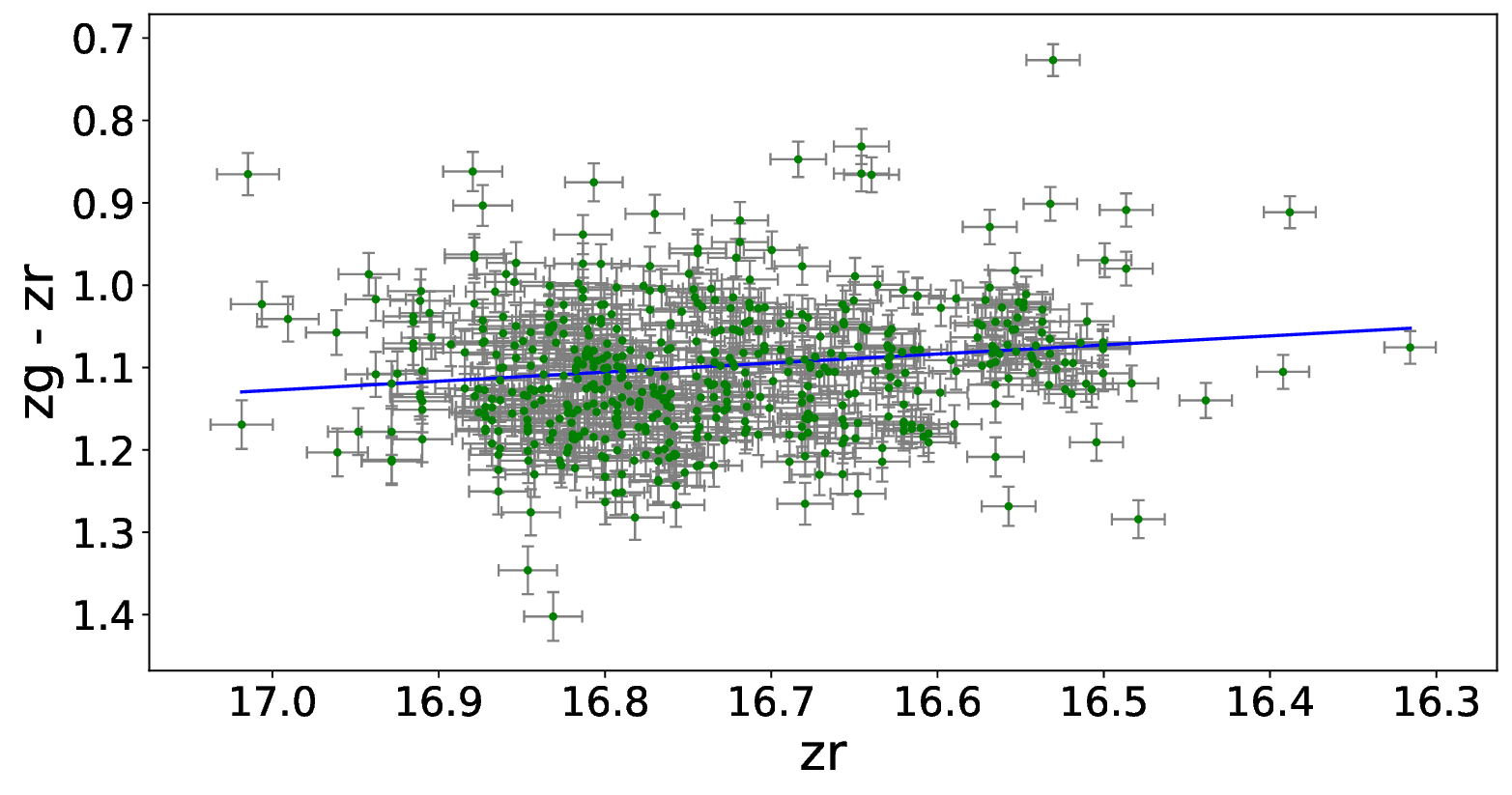}
		\includegraphics[width=0.79\linewidth]{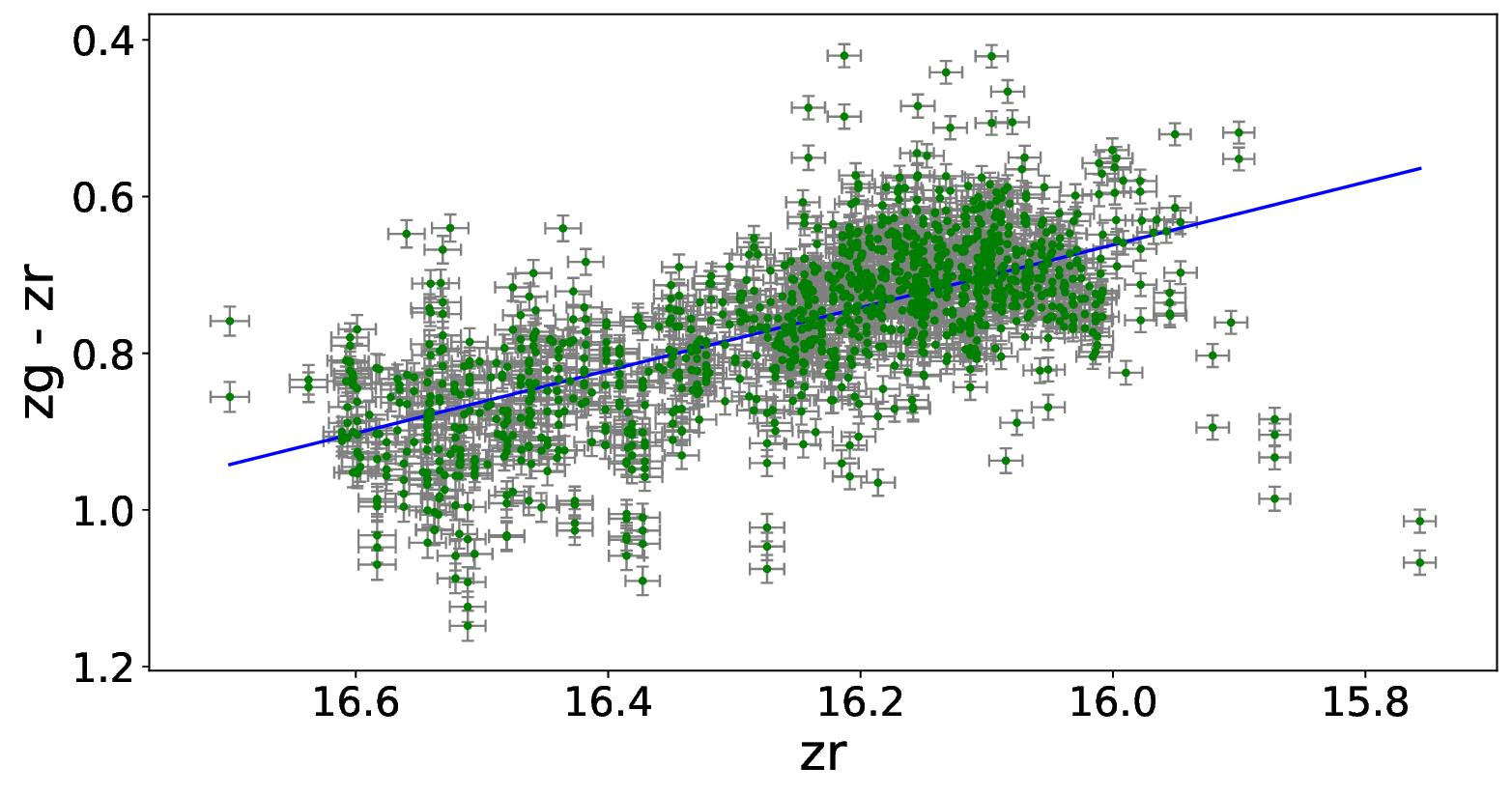}
		\caption{Colour-magnitude diagrams for the four CLAGNs reported
		in this work. From {\it top} to {\it bottom}: J0751+4948, 
		J1020+2437, J1203+6053, and J1344+5126. Their $k$ values 
		determined from a linear fit (solid line in each panel) are 
		respectively 0.376$\pm$0.004, 0.679$\pm$0.008, 0.110$\pm$0.009,
		and 0.401$\pm$0.003. }
		\label{fig:src}
	\end{figure}
	
	\begin{figure*}
		\centering
		\includegraphics[width=0.6\linewidth]{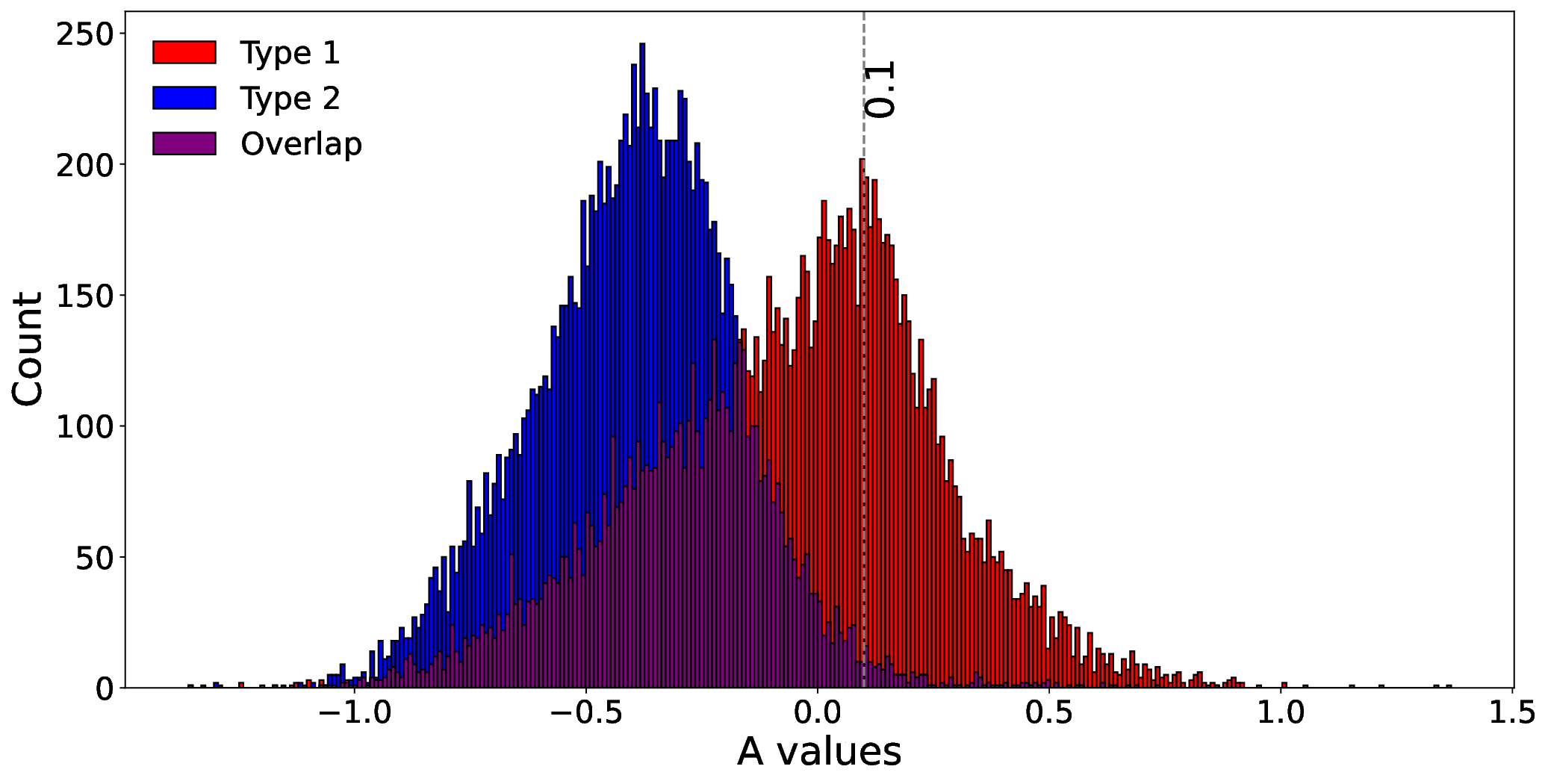}
		\caption{Distributions of the $k$ values (the slopes) of the
		linear fits to each AGN's CM variation data points. 
		A dashed line at $k = 0.1$ is drawn, type 2 AGNs above which
		are checked as potential CLAGN candidates.}
		\label{fig:fit1+2}
	\end{figure*}
	
\subsection{Archival data}
\label{subsec:ad}

Different archival photometric data were used in the target selection 
(Section~\ref{subsec:ts}) and analysis (Section~\ref{sec:ar}). The majority
of data used were from the Zwicky Transient Facility survey 
(ZTF; \citealt{bkg+19}). The magnitude data at its $zg$- and $zr$-band 
for the sources mentioned in this work were obtained. When querying 
the ZTF data, 
we set catflags = 0 and chi $<$ 4 in order to construct clean and high-quality 
light curves for the sources. 

For our targets studied in detail, we also used the $V$-band magnitude
data from the Catalina Real-time Transient Survey (CRTS; \citealt{ddm+09})
and data at the cyan (420--650 nm; $ac$) and orange (560--820 nm; $ao$) bands
from the Asteroid Terrestrial-impact Last Alert System 
(ATLAS; \citealt{tdh+18}). The ATLAS's two wide bands cover the wavelength
ranges of the SDSS's $g+r$ and $r+i$ bands, respectively. In addition, the
mid-infrared (MIR) data obtained from the post-cryogenic phase survey 
of the Wide-field Infrared Survey Explorer (WISE; \citealt{wem+10}) were 
included. The magnitude data are at two bands, W1 (3.4\,$\mu$m) and W2 
(4.6\,$\mu$m).
	
	\subsection{Target selection}
	\label{subsec:ts}

We used the SQL query tool of the SDSS SkyServer\footnote{\url{https://skyserver.sdss.org/dr16/en/tools/search/sql.aspx} to extract spectral data 
from the SDSS's Data Release 16 (DR16; \citealt{dr16}) for sources 
classified as `galaxy' 
with subclass `AGN'. These sources are considered as type 2 AGNs in
the database. In order to} have decent spectra that cover the 
H${\alpha}$ and H${\beta}$ BELs, we further required a median 
signal-to-noise ratio {\tt snmedian} greater than 10 and a redshift ($z$) 
less than 0.5.
In total, we obtained 16,919 spectra 
(without counting multiple spectra of the same sources). For the AGNs,
we selected those with ZTF data points at $zr$-band greater than 43. This 
requirement was set in order to have light curves with sufficient data 
points for
colour-magnitude (CM) analysis, and the number 43 was read from the 
distribution of the ZTF data points for the AGNs. Using the $zg$- and
$zr$-band data, we calculated the colour $zg-zr$, where we required 
the magnitudes at the two bands in the calculation be taken within one day.
We obtained the CM ($zg-zr$ versus $zr$) diagrams for each AGN;
examples of the CM diagrams are shown in Figure \ref{fig:src}. 
We then fit the CM data points with a linear function of 
$zg-zr = k\times zr + c$, where $k$ is the slope and $c$ is a constant.
The distribution of the $k$ values obtained for more than ten thousands type 2 
AGNs is shown in Figure~\ref{fig:fit1+2}. 
We noted that many of the $k$ values were erroneous, 
resulting in unreal BWB patterns, 
because some sporadic variations induced the fitting results. 
In any case, the distribution can serve as a tool for initial source selection.
As a starting point, we chose $k\geq 0.1$ as a threshold and found there were 
177 sources, approximately 1.4\% of the type 2 AGNs shown in
Fig.~\ref{fig:fit1+2} (compared to $\sim$34\% of the type 1 AGNs 
with $k\geq 0.1$ in the figure). After examining the CM diagrams by eye,
we finally found 73 sources with
clear BWB variations. Upon the starting of this work, 13 of the 73 sources 
have been reported as CLAGNs, already affirming the efficiency of our method
in identifying CLAGN candidates from type 2 to type 1. In our 
spectroscopy identification (Section~\ref{sec:obs}), we observed nine
of the 60 candidates. The basic information for the observed sources,
including their maximum optical and MIR magnitude changes ($\Delta zg$,
$\Delta zr$, $\Delta W1$, and $\Delta W2$),
is given in Table~\ref{tab:info1}.\footnote{In calculating $\Delta W1$ and $\Delta W2$, 1--2 obvious outliers of each MIR light curve were excluded, which
could cause erratic results.}


As a test, we also conducted a similar analysis to type 1 AGNs in the SDSS 
database. Sources with spectra classified as `QSO' in the SDSS DR16 
database were selected, while the same criteria as those for the Type 2 AGNs
were applied to the source selection.
In total, we retrieved 19,023 spectra. Among them, we selected those with 
ZTF $zr$ data points greater than 70.
 Their $k$-value distribution is shown in Figure \ref{fig:fit1+2}. 
The idea is that these type 1 AGNs with negative $k$ values would have a 
redder-when-brighter pattern, which is inconsistent with their type. 
This contradiction would suggest that they had the CL from type 1 to type
2. We selected a few sources with $k \leq -0.7$ as targets;
in our samples, there were
$\sim$2.7\% type 2 AGNs and $\sim$8.9\% type 1 AGNs satisfying the selection
condition.
Three of them were observed in last year's observing run.
	
\section{LJT and DOT Observations}
\label{sec:obs}

	\begin{table*}
		\centering
		\caption{Basic informations of all observed sources}
		\label{tab:info1}
		\begin{threeparttable}
			\begin{tabular}{lcccccccc}
				\hline\hline
				Target & R.A. (J2000)  & Decl. (J2000) & $z$ & $\Delta zg$ & $\Delta zr$ & $\Delta W1$ & $\Delta W2$ & k  \\ 
				\hline
				\multicolumn{9}{l}{Confirmed changing-look AGN}\\
				\hline
				J0751+4948 & $07^h$$51^m$$51^s.89$  & +$49^\circ$$48^\prime$$51^{\prime \prime}$.54 & 0.0244 & 1.127$\pm$0.013 & 0.870$\pm$0.011 & 0.940$\pm$0.021 & 1.264$\pm$0.025 & 0.376$\pm$0.004\\
				J1020+2437$^\ast$ & $10^h$$20^m$$38^s.50$  & +$24^\circ$$37^\prime$$08^{\prime \prime}$.35 & 0.1894 & 1.859$\pm$0.048 & 1.227$\pm$0.031 & 1.167$\pm$0.043 & 1.363$\pm$0.069 & 0.679$\pm$0.008\\
				J1150+3503$^\ast$ & $11^h$$50^m$$00^s.57$  & +$35^\circ$$03^\prime$$56^{\prime \prime}$.71 & 0.0611 & 1.102$\pm$0.024 & 1.358$\pm$0.020 & 0.492$\pm$0.035 & 0.922$\pm$0.061 & 0.20$\pm$0.01\\
				J1203+6053 & $12^h$$03^m$$49^s.21$  & +$60^\circ$$53^\prime$$17^{\prime \prime}$.45 & 0.0655 & 0.994$\pm$0.032 & 0.774$\pm$0.018 & 0.514$\pm$0.020 & 0.535$\pm$0.022 & 0.11$\pm$0.01\\
				J1344+5126 & $13^h$$44^m$$19^s.60$  & +$51^\circ$$26^\prime$$24^{\prime \prime}$.66 & 0.0629 & 1.328$\pm$0.021 & 0.970$\pm$0.014 & 0.825$\pm$0.026 & 1.019$\pm$0.035 & 0.401$\pm$0.003\\
				\hline
				\multicolumn{9}{l}{Other candidate}\\
				\hline
				J1053+4929 & $10^h$$53^m$$44^s.13$  & +$49^\circ$$29^\prime$$55^{\prime \prime}$.99 & 0.1404 & 0.892$\pm$0.023 & 0.648$\pm$0.019 & 0.476$\pm$0.031 & 0.618$\pm$0.060 & 0.14$\pm$0.01\\
				J1246$-$0156& $12^h$$46^m$$22^s.70$ & $-01^\circ$$56^\prime$$28^{\prime \prime}$.49 & 0.0844 & 0.894$\pm$0.028 & 0.591$\pm$0.020 & 0.589$\pm$0.037 & 0.955$\pm$0.079 & 0.46$\pm$0.03\\
				J1252+0717 & $12^h$$52^m$$52^s.61$  & +$07^\circ$$17^\prime$$57^{\prime \prime}$.67 & 0.1082 & 0.548$\pm$0.026 & 0.449$\pm$0.021 & 0.778$\pm$0.046 & 1.25$\pm$0.13 & 0.34$\pm$0.04\\
				J1423+2454 & $14^h$$23^m$$52^s.09$  & +$24^\circ$$54^\prime$$17^{\prime \prime}$.14 & 0.0744 & 0.828$\pm$0.025 & 0.564$\pm$0.020 & 0.948$\pm$0.043 & 1.488$\pm$0.085 & 0.679$\pm$0.008\\
				\hline
				\multicolumn{9}{l}{Type 1 AGN}\\
				\hline
				J1127+2654 & $11^h$$27^m$$36^s.38$  & +$26^\circ$$54^\prime$$50^{\prime \prime}$.55 & 0.3792 & 0.160$\pm$0.020 & 0.160$\pm$0.020 & 0.303$\pm$0.044 & 0.345$\pm$0.038 & $-$0.758$\pm$0.001\\
				J1527+2233 & $15^h$$27^m$$57^s.67$  & +$22^\circ$$33^\prime$$04^{\prime \prime}$.02 & 0.2539 & 0.148$\pm$0.016 & 0.148$\pm$0.016 & 0.312$\pm$0.029 & 0.368$\pm$0.038 & $-$0.748$\pm$0.001\\
				J1606+2903 & $16^h$$06^m$$28^s.07$  & +$26^\circ$$29^\prime$$03^{\prime \prime}$.83 & 0.4342 & 0.424$\pm$0.030 & 0.281$\pm$0.023 & - & - & $-$0.865$\pm$0.001\\
				\hline
				
			\end{tabular}   

		\end{threeparttable}       
		\begin{tablenotes}
			\centering
		\item $^\ast$ marks the two sources identified as CLAGN in \citet{wwg+24}. 
		\end{tablenotes}            
	\end{table*}      
	

	\begin{table*}
		\centering
		\caption{Information of spectroscopic observations with LJT and DOT}
		\label{tab:info2}
		\begin{threeparttable}
			\begin{tabular}{lccccc}
				\hline\hline
				Target & Telescope & Date & Exposure & Seeing & Standard  \\ 
				& & & (sec) & (arcsec) \\
				\hline
				\multicolumn{6}{l}{Confirmed changing-look AGN}\\
				\hline
				J0751+4948 & LJT & 2024-03-18 & 600  & 1.5 & BD+33d2642 \\
				J1020+2437 & LJT & 2024-03-18 & 2000 & 1.5 & BD+33d2642 \\
				J1150+3503 & DOT & 2024-03-15 & 1260 & 1.5 & Feige66    \\
				J1203+6053 & DOT & 2024-03-15 & 1440 & 1.5 & Feige66    \\
				J1344+5126 & LJT & 2024-03-17 & 1600 & 1.8 & Feige66    \\
				\hline
				\multicolumn{6}{l}{Other candidate}\\
				\hline
				J1053+4929  & DOT & 2024-03-15 & 1800 & 1.5 & Feige66   \\
				J1246$-$0156& LJT & 2024-03-18 & 1800 & 1.5 & BD+33d2642 \\
				J1252+0717  & LJT & 2024-03-17 & 2000 & 1.9 & Feige66\\
				J1423+2454  & LJT & 2024-03-18 & 2100 & 1.5 & BD+33d2642 \\
				\hline
				\multicolumn{6}{l}{Type 1 AGN}\\
				\hline
				J1127+2654 & LJT & 2024-03-18 & 1500 & 1.6 & BD+33d2642 \\
				J1527+2233 & LJT & 2024-03-18 & 1200 & 1.5 & BD+33d2642 \\
				J1606+2903 & LJT & 2024-03-18 & 900  & 1.6 & BD+33d2642 \\
				\hline
				
			\end{tabular}   
		\end{threeparttable}       
	\end{table*}      

	\subsection{Spectroscopy}

Among the selected AGN sources (Section~\ref{subsec:ts}), 
we chose our targets mainly based on
their visibility and brightnesses; LJT and DOT both have a limiting
magnitude of approximately 19 for spectroscopic observations. The information 
for the
targets and observations is provided in Table~\ref{tab:info2}.
Nine type~2 AGNs and three type~1 AGNs were observed.
For LJT observations, the instrument used was the Yunnan Faint Object 
Spectrograph and Camera 
(YFOSC). This instrument has a 2048$\times$4096\,pixel$^2$ back-illuminated 
Charge-Coupled Device (CCD), with a pixel scale of 0.283 arcsec\,pixel$^{-1}$.
The grism used was G3, which provides a wavelength coverage of 340--910\,nm 
and a spectral dispersion of 0.29\,nm\,pixel$^{-1}$. We chose a long slit
with a width of 2.5\,arcsec in all exposures. 
In addition, spectra for wavelength and flux calibrations were also taken, 
which were those of a He-Ne lamp and a spectrophotometric standard,
respectively. 
	
We also conducted spectroscopic observations of three sources using DOT. 
The instrument was the ARIES-Devasthal Faint Object 
Spectrograph and Camera (AD-FOSC), whose detector is a 4k$\times$4k\,pixel$^2$ 
CCD. For all exposures, we chose the 132R-600\,gr/mm grism, which provides
a spectral dispersion of 0.10\,nm\,pixel$^{-1}$ and a wavelength coverage
of 350--700\,nm. The slit used was 8-arcmin long, with a width of 
2.0\,arcsec. 
Wavelength and flux calibrations were performed by taking the spectra
of Neon and Argon and the spectra of a spectrophotometric standard, 
respectively.
Because the guiding system of the telescope was not functioning, it
was suggested that one exposure
be a maximum of 600\,sec. The exposures
of the three sources given in Table~\ref{tab:info2} consist of 
2--3 $\leq 600$\,sec exposures.
	
	\subsection{Data reduction}
We used the IRAF tasks for data reduction. The spectrum images were bias 
subtracted and flat fielded. Spectra of the sources were extracted, to
which wavelength and flux calibrations were conducted. For the DOT observations,
we obtained the final spectrum of each source by averaging 2--3 spectra,
respectively extracted from the $\leq 600$\,sec exposures.
	
	\section{Analysis and Results}
	\label{sec:ar}
	
	\begin{figure*}
		\centering
		\includegraphics[width=0.49\linewidth]{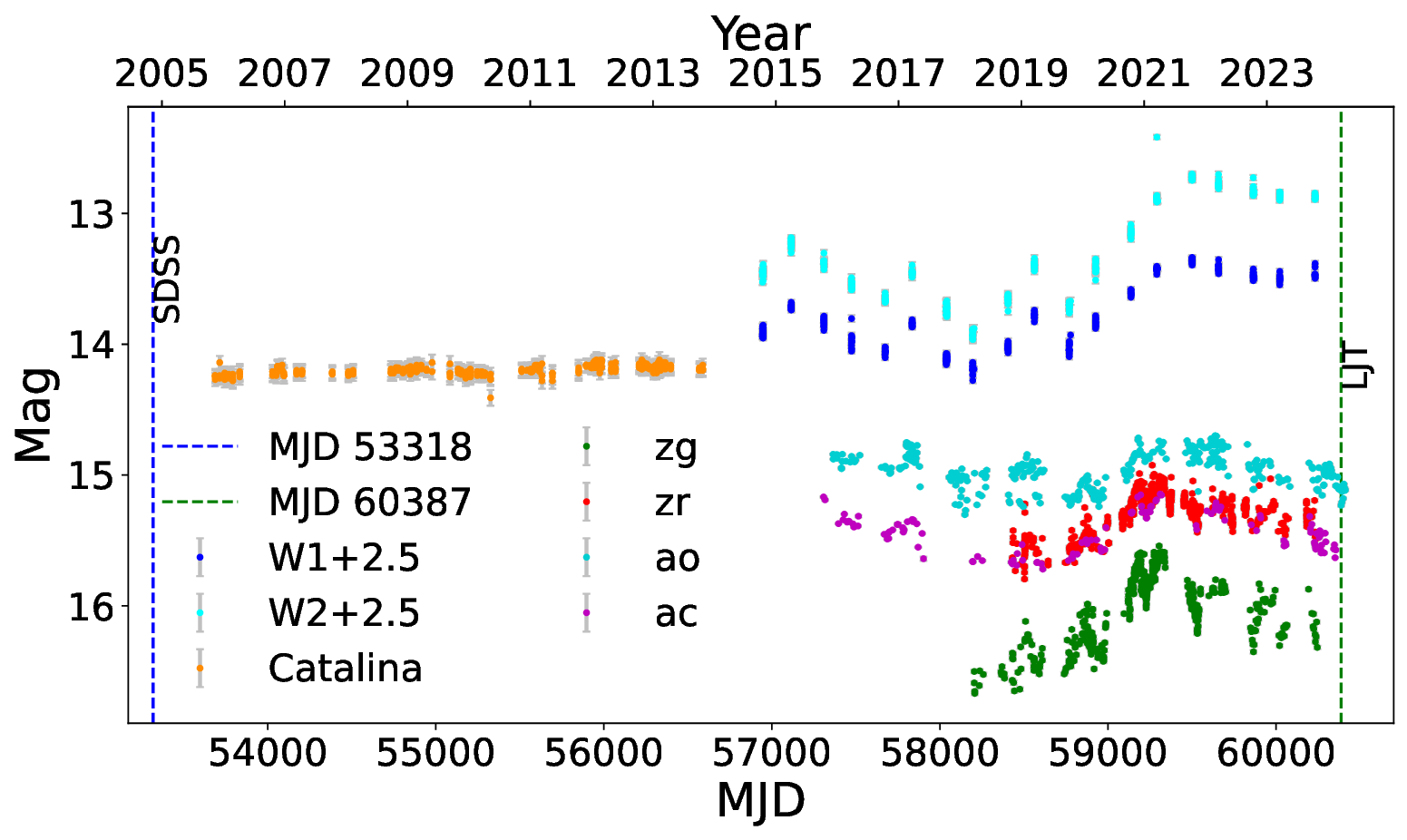}
		\includegraphics[width=0.49\linewidth]{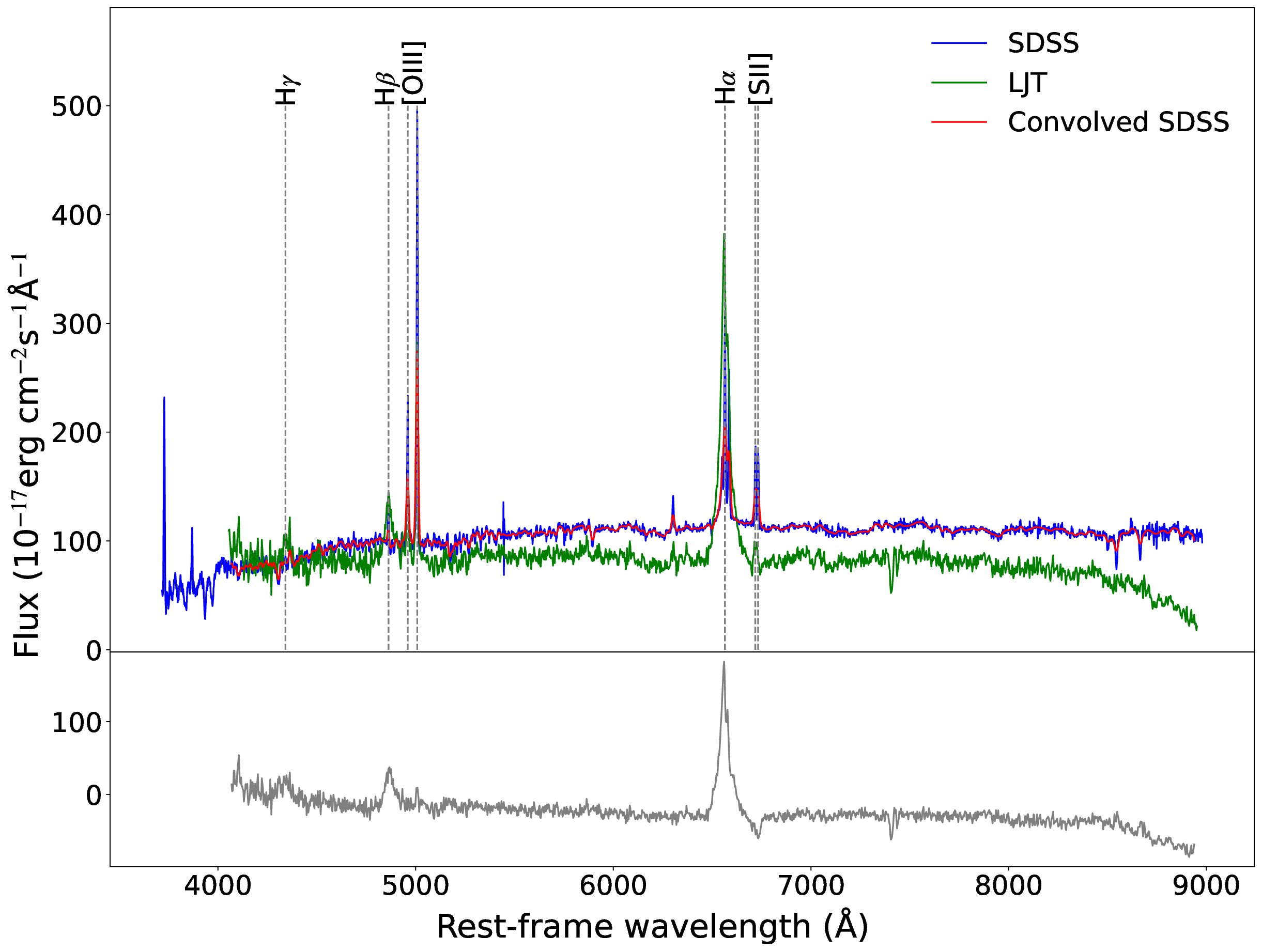}
		\caption{Optical and MIR light curves ({\it left})
		and spectra ({\it right}) of J0751+4948. Two vertical dashed 
		lines in the left panel mark 
		the observation times of the SDSS and LJT spectra shown in
		the right upper panel. 
		The two spectra are vertically shifted for clarity. 
		In the {\it right lower} panel, a difference spectrum between
		the two spectra is shown. }
		\label{fig:src1}
	\end{figure*}

	\begin{figure*}
		\includegraphics[width=0.49\linewidth]{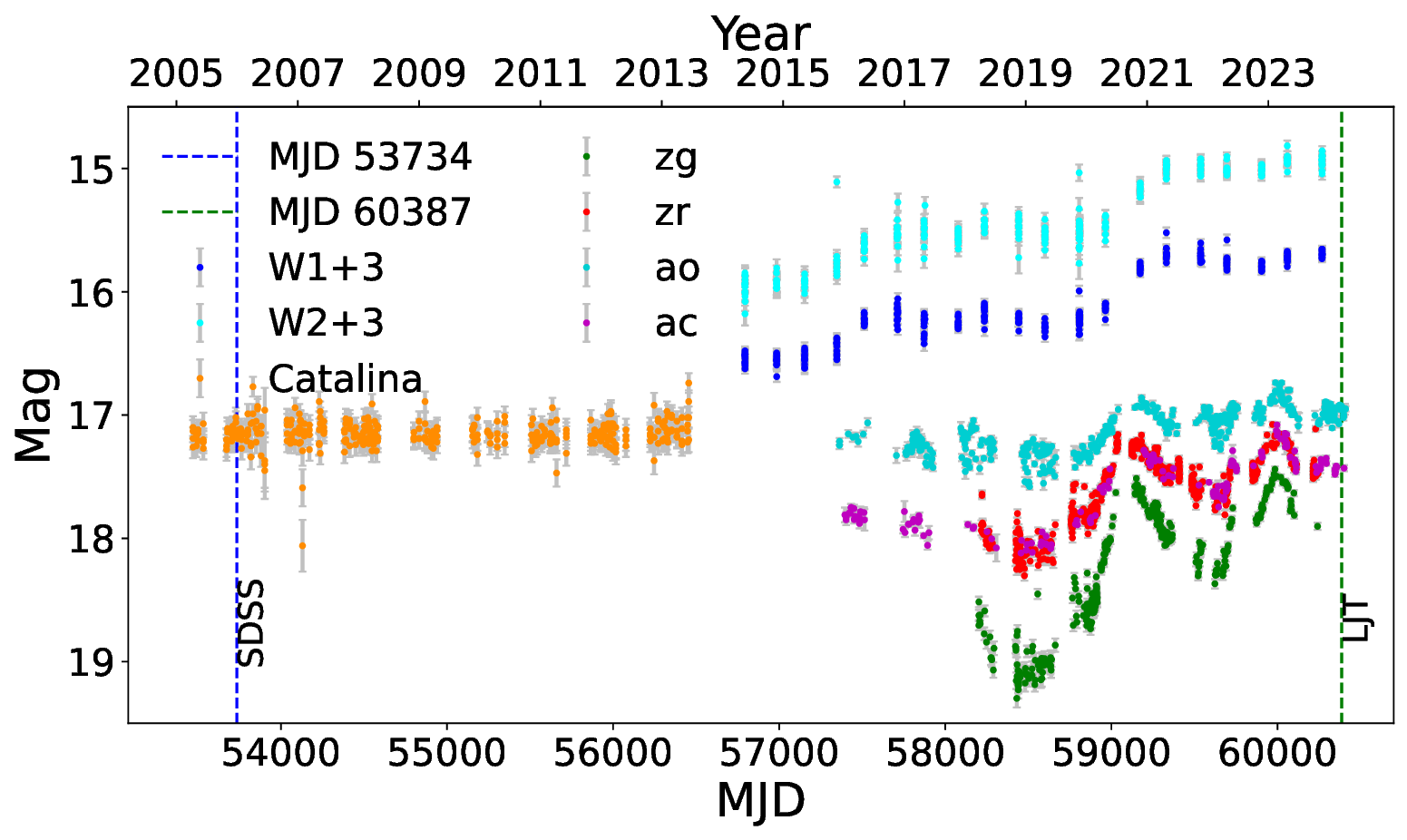}
		\includegraphics[width=0.49\linewidth]{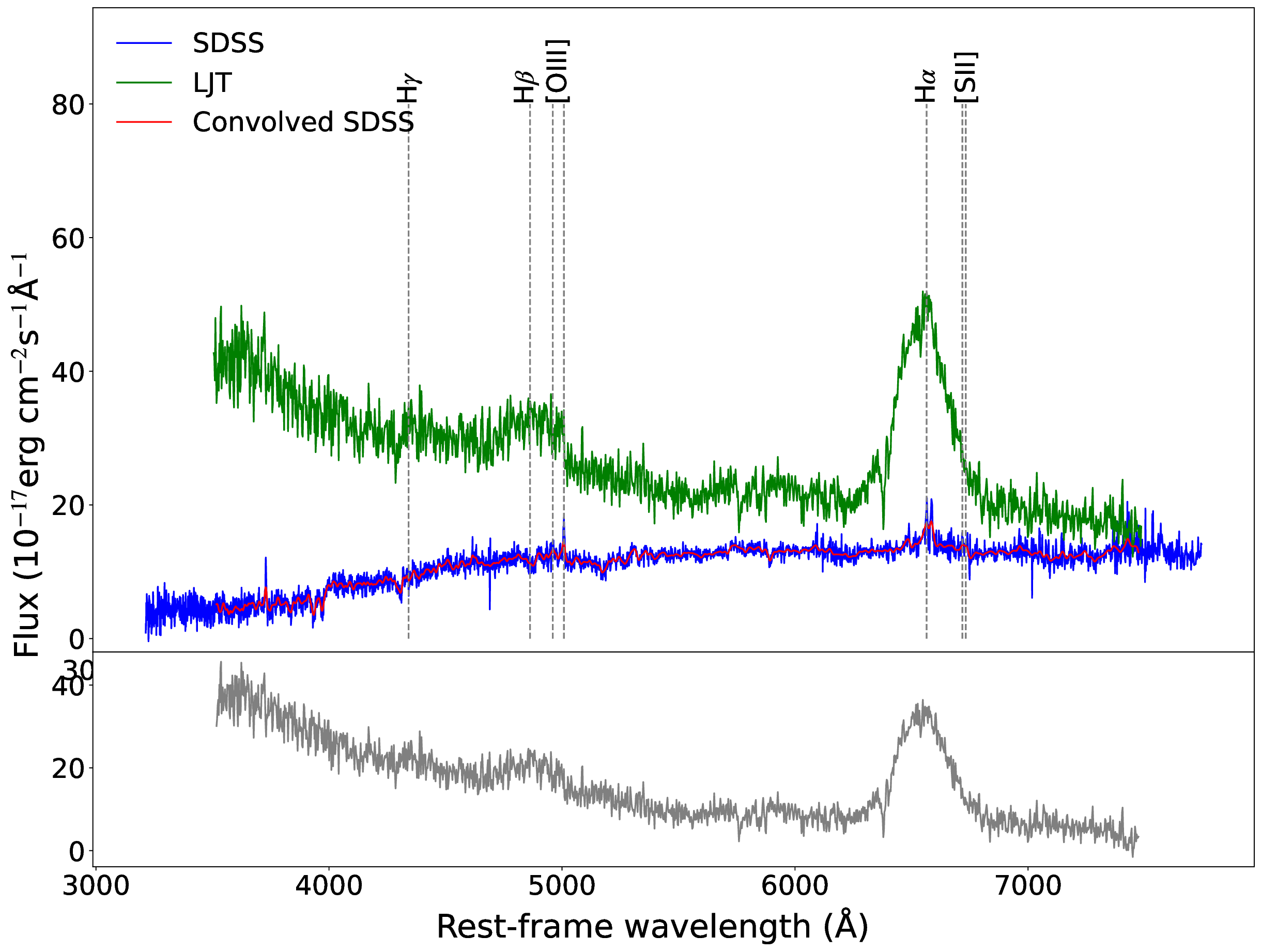}
		\caption{Same as Fig.~\ref{fig:src1} for J1020+2437.}
		\label{fig:src2}
	\end{figure*}	

	\begin{figure*}
		\includegraphics[width=0.49\linewidth]{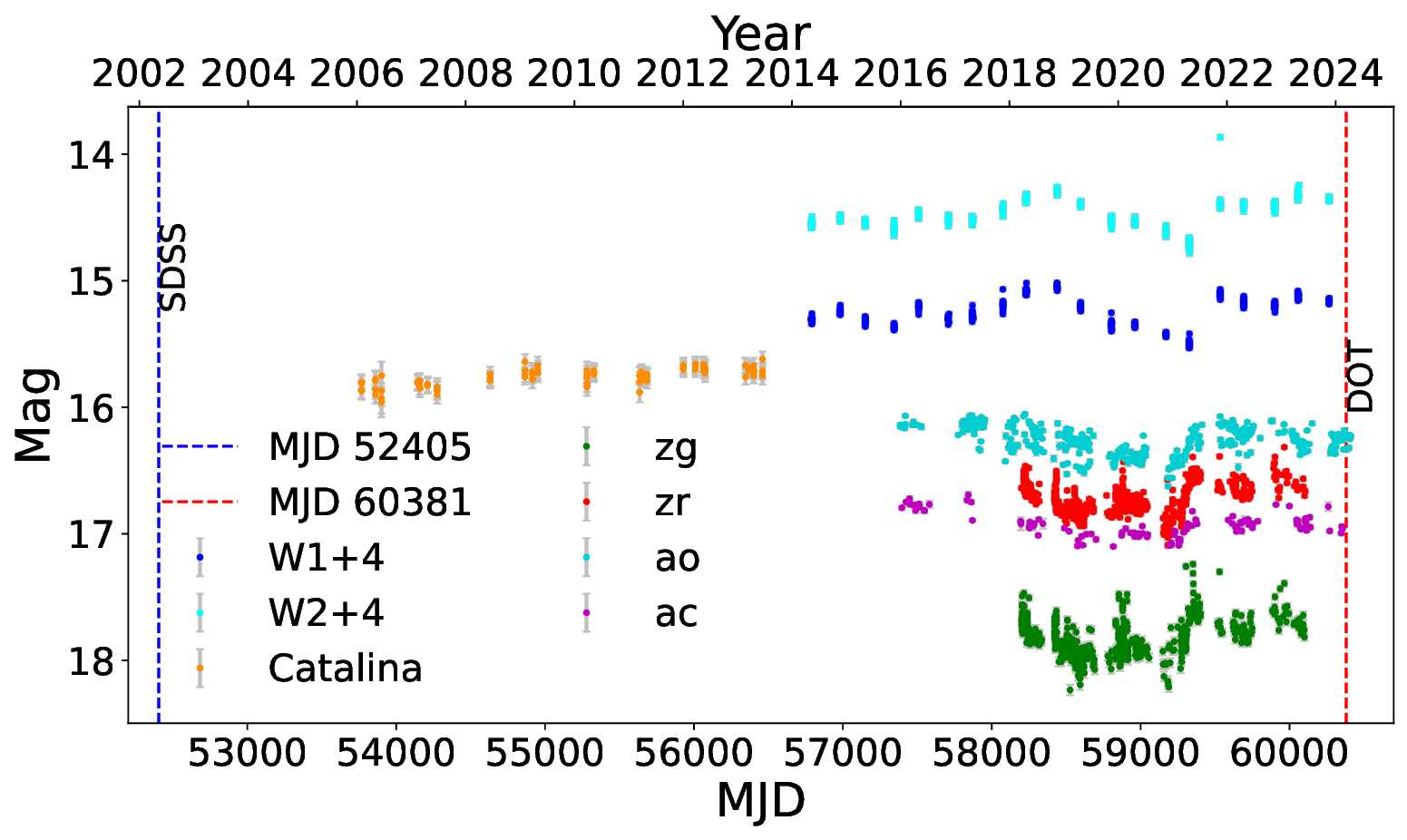}
		\includegraphics[width=0.49\linewidth]{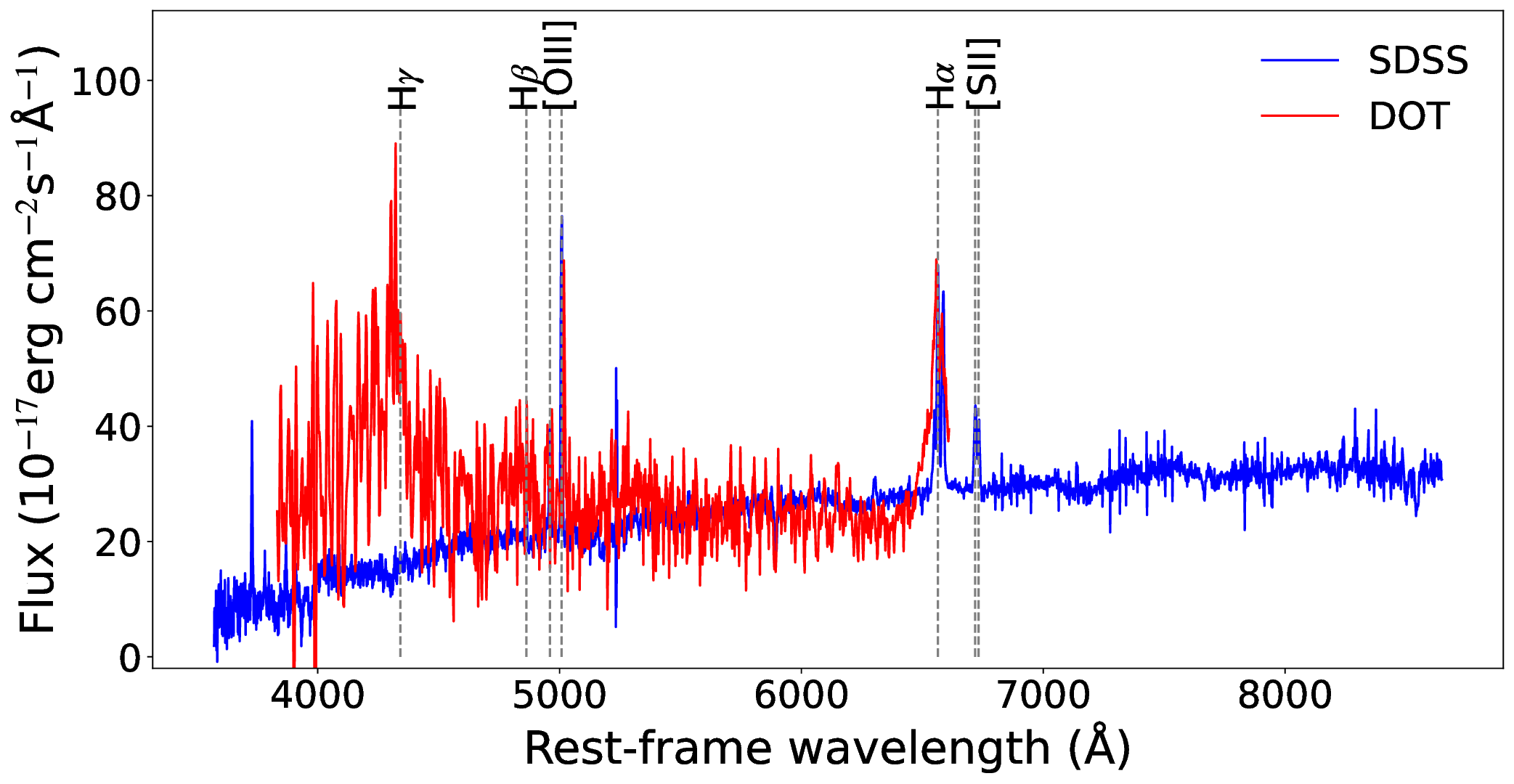}
		\caption{Same as Fig. \ref{fig:src1} for J1203+6053,
		but because the DOT spectrum is 
		noisy, no difference spectrum for it is obtained and shown.}
		\label{fig:src3}
	\end{figure*}

	\begin{figure*}
		\includegraphics[width=0.49\linewidth]{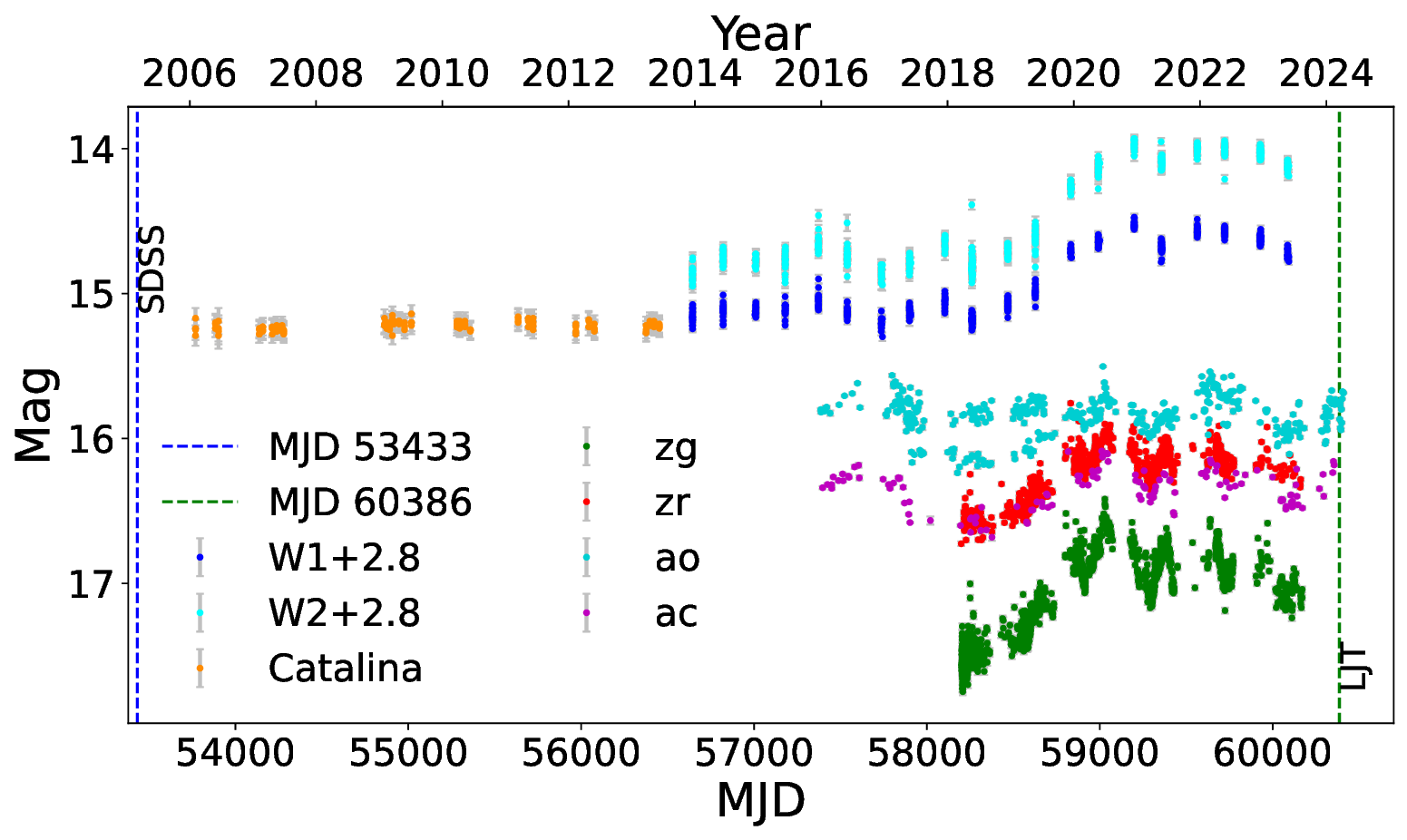}
		\includegraphics[width=0.49\linewidth]{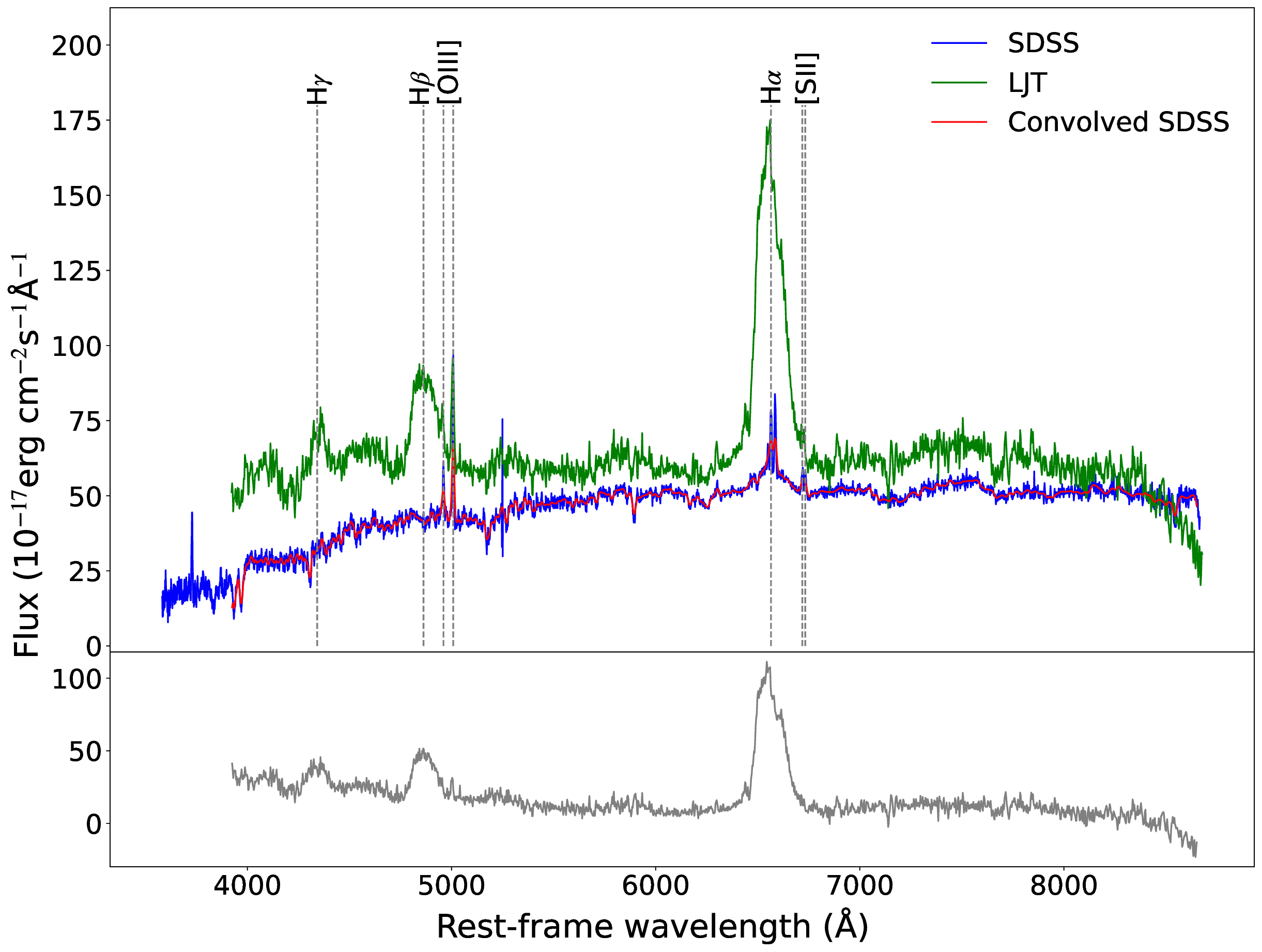}
		\caption{Same as Fig. \ref{fig:src1} for J1344+5126.}
		\label{fig:src4}
	\end{figure*}
	
Among the nine targets selected from type 2 AGNs, three of which, 
J0751+4948, J1020+2437, and J1344+5126, should be type 1.9 based on our 
analysis (Table~\ref{tab:fitting}),
we identified four as CLAGNs from our observations.  
For J1150+3503, its DOT spectrum is of bad 
quality, suffering large uncertainties. However, since its
CL transition has been reported by \citet{wwg+24}, we included
it in Table~\ref{tab:info2}.
We respectively describe the analysis and results below 
in Section~\ref{subsec:cl} \& \ref{subsec:can} for the nine
sources.

The three type 1 AGNs we observed did not show a transition from type 1 
to type 2.
We present analysis of their spectra in Section~\ref{subsec:t1}, and
their CM diagrams are displayed in Fig.~\ref{figa:cm}
in Appendix~\ref{sec:src02}.

To obtain measurements of the prominent emission lines, 
namely H$\alpha$ and 
H$\beta$, for comparison, we employed the PYTHON QSO fitting code 
(PYQSOFIT; \citealt{gsw18}). The full-width at half maximum (FWHM), 
the equivalent width (EW), and the line flux of each of the two lines were 
obtained by fitting them with PYQSOFIT. 
These fitting results, along with the peak wavelength determined for each 
line's broad component, are given in Table~\ref{tab:fitting}.
Details of the spectral fitting are presented in 
Fig.~\ref{fig:sf1}--\ref{fig:sf4} in Appendix~\ref{sec:fit}.

For the measurements given in Table~\ref{tab:fitting}, the systematic 
uncertainties should be considered. We examined each LJT or DOT spectrum and
chose several continuum regions of approximately the same flux level but 
different noise levels. The average fluxes of these chosen regions were 
calculated. We then compared these averages to that of the region with 
the lowest 
noise, and the average of their differences was adopted as the systematic 
uncertainty of a spectrum. Using this method, we estimated uncertainties of
13 per cent, 9 per cent, 15 per cent, and 8 per cent, respectively,
for spectra of J0751+4948, J1020+2437, J1203+6053, and J1344+5126.

We estimated the mass of the black hole (BH) $M_{\rm BH}$ in the CLAGNs
with the following formula from \citet{vp06},
\begin{equation}
	\begin{split}
	\log \left(M_{\mathrm{BH}} / M_{\odot}\right)=  
	\log \left[\left(\frac{\mathrm{FWHM}(\mathrm{H} \beta)}{\mathrm{km}\ \mathrm{s}^{-1}}\right)^{2}\left(\frac{L_{5100}}{10^{44}\ \mathrm{erg}\ \mathrm{s}^{-1}}\right)^{0.5}\right] \\
		+0.91,
	\end{split}
	\label{equ:bm}
\end{equation}
where $L_{5100}$ is the luminosity at 5100\,\AA.
We also estimated the Eddington ratio $\lambda_{\text{Edd}}$ 
($=L_\text{bol}/L_{\text{Edd}}$) for the accretion of a BH. 
For $z<0.8$, $L_{\text{bol}} = 9.26 \times {L_{5100}}$, and
$L_{\text{Edd}} = 1.38 \times 10^{38} M_{\text{BH}}/M_{\odot} $\,erg\,s$^{-1}$ 
\citep{rls+06}. 
In the estimation, the above systematic uncertainties were included.
	
	\subsection{Changing-look AGNs}
	\label{subsec:cl}
	
	\subsubsection{J0751+4948}

The optical and MIR light curves are shown in  
Fig.~\ref{fig:src1}. A flux rise can be observed since approximately
MJD~58200, with the rise in $zg$ appearing faster. 
The $\Delta$$zg$ and $\Delta$W2 values between the start of the rise 
and the variation peak (after MJD~59000) are $\simeq$1.1 and $\simeq$1.3, 
respectively.
These variation features (cf., Fig.~\ref{fig:src}) of the source made it 
selected 
by us. The LJT spectrum, compared to the SDSS spectrum taken $\sim$19\,yr ago,
shows broader and stronger \ha\ emission. It also shows the emergence of 
an \hb\ line, maybe an H${\gamma}$ line as well. To illustrate the differences, 
we subtracted the
SDSS spectrum convolved with the LJT's spectral resolution from the LJT 
spectrum, and obtained a difference spectrum. This difference spectrum is shown
in Fig.~\ref{fig:src1}, and the changes in \ha\ and \hb\ 
are clearly visible.

The FWHM of H$\beta$ was $\simeq$3370 km\,s$^{-1}$ 
(Table~\ref{tab:fitting}). Using it, we 
obtained $M_{\rm BH}\sim 10^{6.69} M_{\odot}$ from Eq.~\ref{equ:bm}.
The estimated Eddington ratio $\log \lambda_{\text{Edd}}$ was $\sim -2.1$, 
indicating that its accretion mode was more likely 
the ADAF rather than the standard accretion disc flow,
given the threshold of $\lambda_{\text{Edd}}\sim 0.01$ \citep{ss73, nd18}.
	
	\subsubsection{J1020+2437}

Similar to J0751+4948, the source's $zg$ band started brightening faster from 
MJD~58500 before reaching a maximum change of $-$1.4 mag in less than a 
thousand days
(Fig. \ref{fig:src2}).  A notable feature is that the MIR emission has been 
increasing since the beginning of the WISE data. Compared to the SDSS
spectrum taken $\sim$18\,yr ago, our LJT spectrum shows 
a very strong \ha\ line, accompanied with the emergence of
a broad \hb\ component. The FWHM of the strong H$\alpha$ line was
over 10,000\,km\,s$^{-1}$, while a similarly broad but weak \ha\ component 
was required in our fitting of the SDSS spectrum. These results are consistent 
with that reported in \citet{wwg+24} for this source. 
The mass of the BH, estimated from the FWHM of the H$\beta$ broad component, 
was $\sim 10^{8.88} M_{\odot}$. The estimated Eddington ratio 
$\log\lambda_{\text{Edd}}$ was $\sim -1.8$.
	
	\subsubsection{J1203+6053}

Compared to the other sources, flux variations of this source are more
like that of a flickering type, with no major brightening event. 
Carefully examining
the multi-band light curves, there seemed to be a sudden increase starting
from $\sim$MJD~59200 in the optical, possibly accompanied by a delayed
jump in the MIR bands (Fig.~\ref{fig:src3}). The DOT spectrum suffers large 
uncertainties and only captures half of the \ha\ line. We did not obtain
a difference spectrum for this source. In any case, the presence of broad
H$\alpha$ and \hb\ (as well as H$\gamma$) components, as opposed to their
absence in the SDSS spectra, suggests this is a CLAGN.
The BH mass was estimated to be $\sim 10^{7.97} M_{\odot}$,
although this value is highly uncertain due to the limited quality of 
the spectrum. The estimated Eddington ratio $\log \lambda_{\text{Edd}}$ was 
$\sim -1.97$.
	
	\subsubsection{J1344+5126}

This source, similar to J1020+2437, exhibited a $<$1000\,day long optical flux
increase (peaking around MJD~59000), a long-term brightening in the MIR bands, 
and significant spectral changes (Fig.~\ref{fig:src4}). For the latter, 
our LJT spectrum 
shows not only the appearances of strong and broad \ha\ and \hb\ lines, 
but also the certain appearance of an H$\gamma$ line. The turn-on of
a H$\gamma$ line has not been commonly seen in AGNs exhibiting the
H$\beta$ turn-on phenomenon. In addition, some weak absorption features
were possibly detected in the SDSS spectrum, suggesting a relatively strong
contribution from the host galaxy to the observed emission at the time. 
The BH mass was $\sim 10^{7.55} M_{\odot}$,
and the Eddington ratio $\log \lambda_{\text{Edd}}$ was $\sim -2.4$.
Further investigation of this source's general properties is warranted
in order to draw a full picture of this AGN's dramatic activity. 
	
	\subsection{Other targets}
	\label{subsec:can}
We did not detect CL transitions from type 2 to type 1 in the other
four targets selected from the BWB pattern. No significant variations of
the emission lines were seen in the spectra we obtained, as compared to 
the respective SDSS spectra. In the Appendix Fig.~\ref{figa:four}, we show 
their light curves. Comparing the light curves to those of CLAGNs in
Section~\ref{subsec:cl}, we also see
larger $zg$ (or bluer) flux variations, but the amplitudes are smaller.
For example, all $\Delta zg$ (and $\Delta zr$) values are less 
than 0.9\,mag.
In addition, the MIR light curves are either 
relatively flat (as in J1053+4929) or have been decaying recently (as
in the other three sources). These differences will be considered in our 
follow-up work when selecting targets as CLAGN candidates. The further 
selection consideration will then be tested.
	
	\begin{figure}
		\centering
		\includegraphics[width=0.79\linewidth]{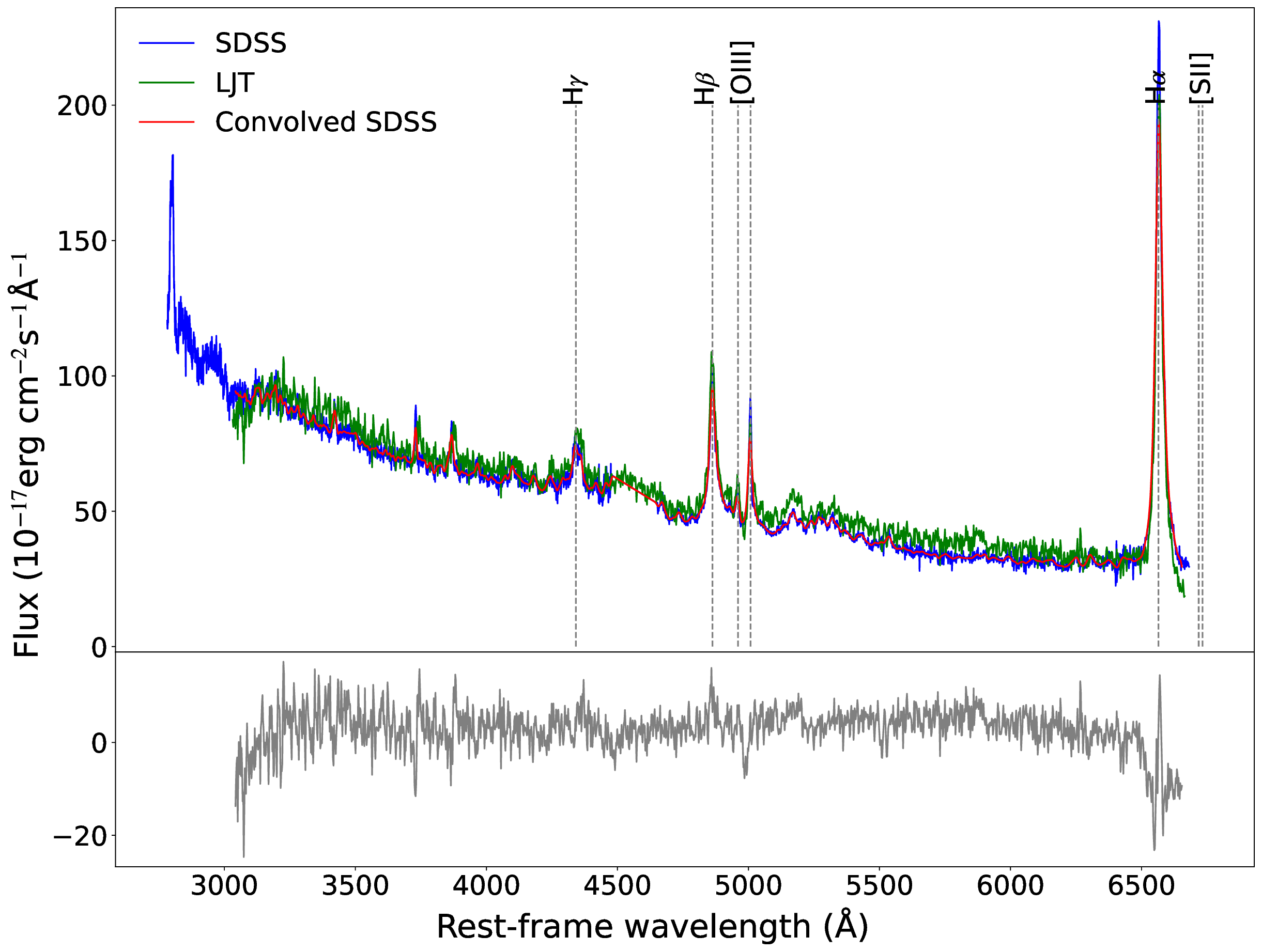}
		\includegraphics[width=0.79\linewidth]{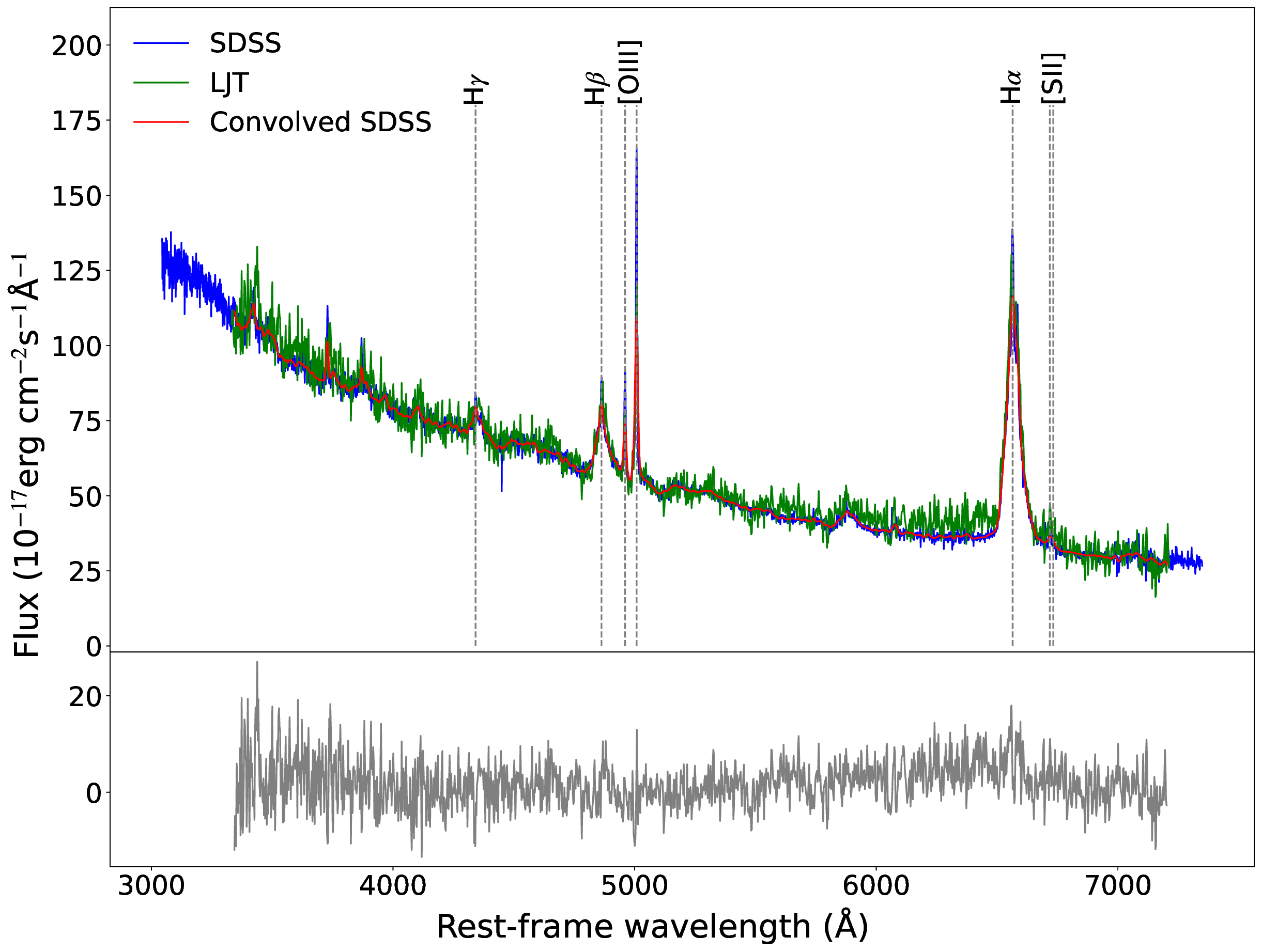}
		\includegraphics[width=0.79\linewidth]{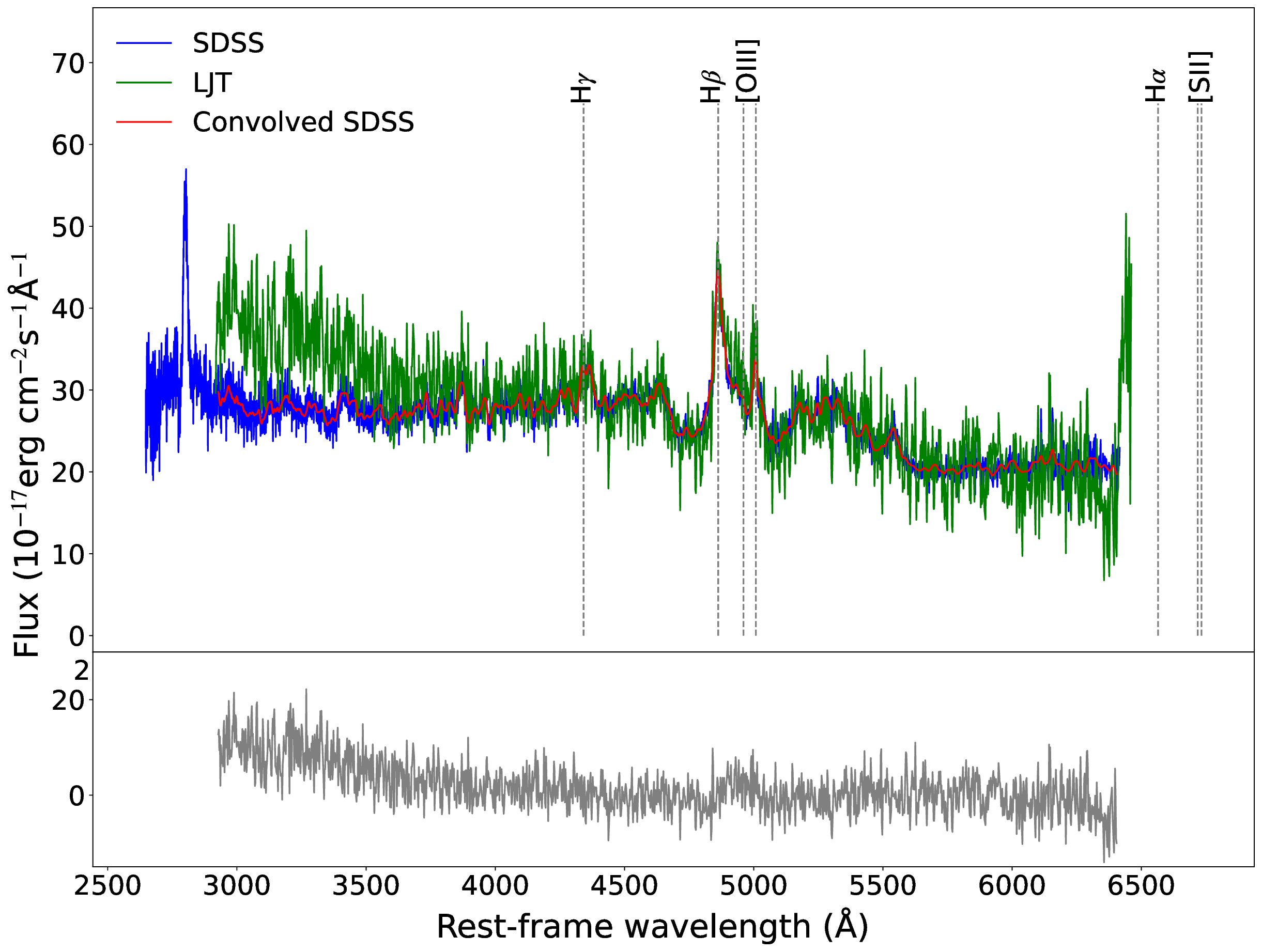}
		\caption{Spectra of three type 1 AGNs we tested, displayed from {\it top} to {\it bottom}: J1127+2654, J1527+2233, and J1606+2903. For each
		source, a difference spectrum is made and shown.}
		\label{fig:src51}
	\end{figure} 
	
	
	\begin{table*}
		\centering
		\caption{Measurements of \ha\ and \hb\ lines from the fitting with PYQSOFIT for the four CLAGNs}
		\label{tab:fitting}
		\begin{tabular}{ccccc}
			\hline
			Line   & (J0751) SDSS   & LJT  & (J1020) SDSS   & LJT      \\ 
			\hline
			
			\multicolumn{5}{l}{H$\alpha$\quad broad}\\
			\hline 
			FWHM     & 3350$\pm$170     & 3858$\pm$66       & 9500$\pm$730    & 10750$\pm$270   \\
			EW       & 218.5$\pm$4.7    & 11620$\pm$29      & 142$\pm$15      & 362.2$\pm$5.2   \\
			Flux     & 1773$\pm$38      & 12670$\pm$320     & 319$\pm$35      & 8860$\pm$130     \\
			Peak     & 6562.21$\pm$0.87 & 6563.65$\pm$0.74  & 6547.9$\pm$5.9 & 6540.7$\pm$3.3  \\
			\hline
			
			\multicolumn{5}{l}{H$\alpha$\quad narrow}\\
			\hline
			FWHM & 302.6$\pm$3.6 & 766$\pm$27   & 354$\pm$16     & 0   \\
			EW   & 161.1$\pm$2.0     & 310$\pm$13   &  24.4$\pm$1.4  & 0   \\
			Flux & 1306$\pm$16   & 3380$\pm$140 & 54.5$\pm$3.1   & 0   \\
			\hline
			
			\multicolumn{5}{l}{H$\beta$\quad broad}\\
			\hline
			FWHM  & 0 & 3370$\pm$240  & 0 & 8720$\pm$780 \\
			EW    & 0 & 190$\pm$17    & 0 & 46.1$\pm$5.5 \\
			Flux  & 0 & 3080$\pm$270  & 0 & 1430$\pm$170 \\
			\hline
			
			\multicolumn{5}{l}{H$\beta$\quad narrow}\\
			\hline
			FWHM & 353.8$\pm$1.9   & 752$\pm$53     & 0 & 0 \\
			EW   & 22.50$\pm$0.79  & 21.6$\pm$9.5   & 0 & 0 \\
			Flux & 273.7$\pm$9.6   & 350$\pm$150    & 0 & 0 \\
			\hline
			
			MJD  & 53318 & 60387 & 53734 & 60387  \\
			\hline
			
			log($M_{\rm BH}$/$M_{\odot}$)& ... &  $6.96^{+0.17}_{-0.20}$ & ... & $8.88^{+0.13}_{-0.14}$  \\
			\hline        
			log($\lambda_{\rm Edd}$)&...&$-$2.11$\pm$0.18&...&$-$1.84$\pm$0.15\\
			\hline
		\end{tabular}\\
		
		\begin{tabular}{ccccc}
			\hline
			Line  & (J1203) SDSS   & DOT  & (J1344) SDSS   & LJT  \\ 
			\hline
			
			\multicolumn{5}{l}{H$\alpha$\quad broad}\\
			\hline 
			FWHM     & 0  & 8020$\pm$240   & 7650$\pm$220   & 8870$\pm$570  \\
			EW       & 0  & 325.1$\pm$9.1  & 371$\pm$12     & 1141$\pm$43    \\
			Flux     & 0  & 5780$\pm$130   & 1122$\pm$36    & 1728$\pm$64  \\
			Peak     & 0  & 6575.4$\pm$2.1 & 6555.7$\pm$2.5 & 6551.1$\pm$2.2 \\
			\hline
			
			\multicolumn{5}{l}{H$\alpha$\quad narrow}\\
			\hline
			FWHM & 430.5$\pm$7.7 & 610$\pm$21     & 299.5$\pm$8.2 & 604$\pm$75  \\
			EW   & 155.5$\pm$3.1 & 19.37$\pm$0.94 & 46.6$\pm$2.6  & 46.6$\pm$6.9   \\
			Flux & 382.9$\pm$7.6 & 361$\pm$18     & 140.4$\pm$7.9 & 72$\pm$11  \\
			\hline
			
			\multicolumn{5}{l}{H$\beta$\quad broad}\\
			\hline
			FWHM  & 0 & 5593.9$\pm$4.5 & 0 & 5599.6$\pm$5.1 \\
			EW    & 0 & 33.7$\pm$1.7   & 0 & 87.2$\pm$3.2    \\
			Flux  & 0 & 874$\pm$44     & 0 & 390$\pm$15 \\
			\hline
			
			\multicolumn{5}{l}{H$\beta$\quad narrow}\\
			\hline
			FWHM & 450.8$\pm$7.5 & 0  & 357.8$\pm$7.6 & 820$\pm$130 \\
			EW   & 10.1$\pm$1.2  & 0  & 4.5$\pm$1.1   & 9.4$\pm$3.1     \\
			Flux & 41.8$\pm$4.9  & 0  & 16.8$\pm$4.0  & 41$\pm$13  \\
			\hline
			
			MJD  & 52405 & 60381 & 53433 & 60386 \\
			\hline
			
			log($M_{\rm BH}$/$M_{\odot}$) & ...     & $7.97^{+0.16}_{-0.18}$ &... & $7.55^{+0.16}_{-0.15}$  \\
			\hline    
			log($\lambda_{\rm Edd}$)&...&$-$1.97$\pm$0.20&...&$-$2.41$\pm$0.10\\
			\hline    
		\end{tabular}\\
		\begin{tablenotes}
			\centering
		\item FWHM, EW, flux, and peak (wavelength) are in units of km\,s$^{-1}$, angstrom (\AA), $10^{-17}$ erg\,cm$^{-2}$\,s$^{-1}$, and angstrom (\AA), respectively. 
		\end{tablenotes}            
	\end{table*}

	\subsection{Type 1 AGNs}
	\label{subsec:t1}
We also tested to select type 1 AGNs with negative $k$ values as the targets,
and to check if they would have a transition from type 1 to type 2 based 
on the CM
property. In addition, their $zr$-band fluctuations were small, with
an amplitude of $\sim$0.2 mag (Fig.~\ref{figa:cm}). However, the LJT spectra
did not show any significant spectral changes compared to the respective
SDSS spectra (Fig. \ref{fig:src51}).
These cases may indicate that AGNs could stay stable, without showing any
significant variations, for a long time period of $\sim$2000\,day. Further
consideration to improve our selection for the turn-off transition will be
taken.

\section{Discussion and Summary}
	\label{sec:ds}
As suggested in \citet{zlw+24}, the BWB pattern seen in CM diagrams
of type 2 AGNs may be used to find CLAGNs with transitions from type 2 to 
type 1.
We thus carried out a test observation run and observed nine selected targets.
Among them, two have already been identified as CLAGNs in \citet{wwg+24}, 
and three are newly discovered by us.
The success rate is greater than 50 per cent if we only 
consider our small sample. Comparing the light curves of the sample,
the apparent differences between CLAGNs and non-CLAGNs are that the latter
had smaller magnitude changes ($< 0.9$\,mag in $zg$; 
Table~\ref{tab:info1}) and most of them also had
decaying MIR emissions in recent years (Fig.~\ref{figa:four}). By contrast,
the CLAGNs all had $> -$1.0\,mag flux increases in $zg$, which
were mostly due to a brightening flare-like event
(except J1203+6053). Such events are not seen in the non-CLAGNs
(Fig.~\ref{figa:four}); they instead showed flicker-like variations. 
Moreover,
the CLAGNs showed accompanying MIR brightening. We note that among
the four CLAGNs we observed, J0751+4948 and J1344+5126 had $W1-W2$ colour 
changes from $<$0.5 (galaxy-like) to $>$0.5 (AGN-like; e.g., \citealt{lsp+23})
when they entered their flare-like brightening phase,
while the other two sources 
had the colours always $>$0.5. The MIR activity and the related colour changes 
could be a critical indicator to reflect the optical variations (and thus 
the accretion rate changes; \citealt{swj+17}) 
and to be applied in finding CLAGNs \citep{swj+20}.  
Given the differences, which can be verified from 
observations of more sources, in addition to the simple BWB pattern,
	factors such as optical and MIR
magnitude changes, as well as the association with a major brightening 
event should be considered in our selection method. Hopefully with more 
observations, we would be
able to establish the criteria for more effectively selecting CLAGN candidates.
We note that the
results, in-turn,
would allow us to configure the general properties of CLAGNs.

Our method is similar to those focusing on different aspects of
AGNs. 
Besides the mentioned MIR-variation method, for example, \citet{wwg+24} were able to successfully find CLAGNs among sources showing a mismatch between 
variabilities and previously-classified types. 
Our test to find type 1 to type 2 transitions follow the same idea.
However, the failures (although only with three of the observed
sources) suggests more factors
should be included in addition to the minimum variations in our selection.
\citealt{lmb+22} employed a machine-learning classification tool to
select type 1 AGNs among previously classified type 2 ones, 
and had a success rate of $\geq$ 66 per cent in finding CLAGNs. The 
classification tool should have considered all aspects of AGNs,
including optical variations and related CM behaviours as mentioned
in our method. In comparison, our method is simple and probably more 
direct. 
Physically, large BWB slopes caused by signficant flux changes (for
example, $\Delta zg > -$1.0\,mag in our cases)
likely indicate strong variations due to significant accretion rate changes;
the colour variability and magnitude variability in the optical and
MIR are more likely to exclude the variable obscuration scenario (see, e.g.,
\citealt{ywf+18} for detailed discussion).
We are planning a more complete study by carrying out spectroscopy
of a large sample. We will focus more on whether there is a flare-like
event in association with the BWB variations. The results will possibly 
allow us to refine the selection 
criteria and establish some characteristics of CLAGNs for their 
variability aspect.
	
It has been summarized from analyses of large samples of CLAGNs that
their Eddington ratios tend to be around $10^{-2}$, in a range 
of $\log\lambda_{\rm Edd}$ from $-2.5$ to $-1.0$ \citep{zte+24,ps24}.
The four CLAGNs observed by us all had $\log\lambda_{\rm Edd}$ in this range.
As $\lambda_{\rm Edd}\sim 10^{-2}$ is a transition point for the accretion
mode from the ADAF to the standard thin disc, the $\lambda_{\rm Edd}$ properties
could indicate that the CL phenomenon is caused by the mode transition,
probably in the inner region close to the BH in order to match the
short CL timescales \citep{nd18}.
	
Among the four CLAGNs observed by us, J1344+5126 had the lowest 
$\log\lambda_{\rm Edd}$ value ($\sim -2.4$). It can be noted that in its
SDSS spectrum, weak absorption features were present. 
CLAGNs with similar spectra were detected in studies
such as reported by \citet{mrl+16}, \citet{rac+16} and \citet{ywf+18}.
This type of spectra suggests a weak AGN emission component in the sources
at the time.
In fact, based on our fitting to the SDSS spectra of the four CLAGNs
(Fig.~\ref{fig:sf1}--\ref{fig:sf4}), the host-galaxy emission components were 
strong or dominant. For such cases,
J. Li et al.  (in preparation) have conducted simulation calculations,
and according to their study, the cause of the apparent CL 
phenomenon could be due to significant short-term changes 
in the extreme ultraviolet (EUV)
radiation of an AGN, which is part of the overall flux fluctuations
caused by disc temperature turbulences \citep{cai+18}. The EUV radiation
affects the strengths of
the BELs. At the weak phase of the EUV radiation, the influence of the host 
galaxy often results in characteristics typical of type 1.8/1.9 or even type 2, 
yet intrinsically, the BELs are consistently present.
In other words, the CLAGNs have always been type 1. We note that 
three CLAGNs in our cases (except J1203+6053) had a broad H$\alpha$ component
in their SDSS spectra, and they were mostly type 1.9, not pure type 2.  
The upward brightening fluctuation naturally has a BWB pattern, which 
accompanies the `turn-on' of the BELs.
Thus, this strongly varied EUV-radiation scenario may provide an alternative
explanation for the CL phenomenon. 
	
As a summary, we employed the CM patterns, suggested in \citet{zlw+24}, 
for selecting CLAGN candidates among previously-classified type 2 AGNs
in the SDSS.  
We observed nine candidates, four of which were confirmed to display 
the CL phenomenon in our observations and one of which was identified by 
\citet{wwg+24} as the spectrum we obtained from it was of bad quality.
We also tested to extend the selection
method to previously-classified type 1 AGNs, but none of the three observed 
sources showed a transition to type 2. The results prove that this rather 
simple method can effectively find CLAGNs, while the failure cases suggest the
selection criteria could be refined. We plan to carry out observations
of a large number of possible candidates, which aim to not only identify
CLAGNs, but also draw lines on the properties of the optical/MIR variations 
and related CM changes
between CLAGNs and non-CLAGNs. The program will possibly provide quantified
characteristics for CLAGNs, and help us gain a full understanding of 
this particular phenomenon.
	
\section*{Acknowledgements}

This work was
based on observations obtained with the Samuel Oschin Telescope 48-inch and
the 60-inch Telescope at the Palomar Observatory as part of the Zwicky
Transient Facility project. ZTF is supported by the National Science
Foundation under Grant No. AST-2034437 and a collaboration including Caltech,
IPAC, the Weizmann Institute for Science, the Oskar Klein Center at
Stockholm University, the University of Maryland, Deutsches
Elektronen-Synchrotron and Humboldt University, the TANGO Consortium of
Taiwan, the University of Wisconsin at Milwaukee, Trinity College Dublin,
Lawrence Livermore National Laboratories, and IN2P3, France. Operations are
conducted by COO, IPAC, and UW.

This work made use of data products from the Wide-field Infrared Survey
Explorer, which is a joint project of the University of California, Los
Angeles, and the Jet Propulsion Laboratory/California Institute of Technology,
funded by the National Aeronautics and Space Administration.

We thank the anonymous referee for detailed insightful comments, which
greatly helped improve the manuscript.
This research is supported by the Basic Research Program of Yunnan Province
No. 202201AS070005, the National Natural Science Foundation of China
(12273033), and the Original Innovation Program of the Chinese Academy of 
Sciences (E085021002). L.Z. acknowledges
the support of the science research program for graduate students of Yunnan 
University (KC-24249083).

	\section*{Data Availability}
	The data underlying this article will be shared on reasonable request to the corresponding author.
	
	\bibliographystyle{mnras}
	\bibliography{cl}

	\appendix
	
\section{Colour-magnitude Diagrams of Three Type 1 AGNs}
	\label{sec:src02}
	\begin{figure}
		\centering
		\includegraphics[width=0.79\linewidth]{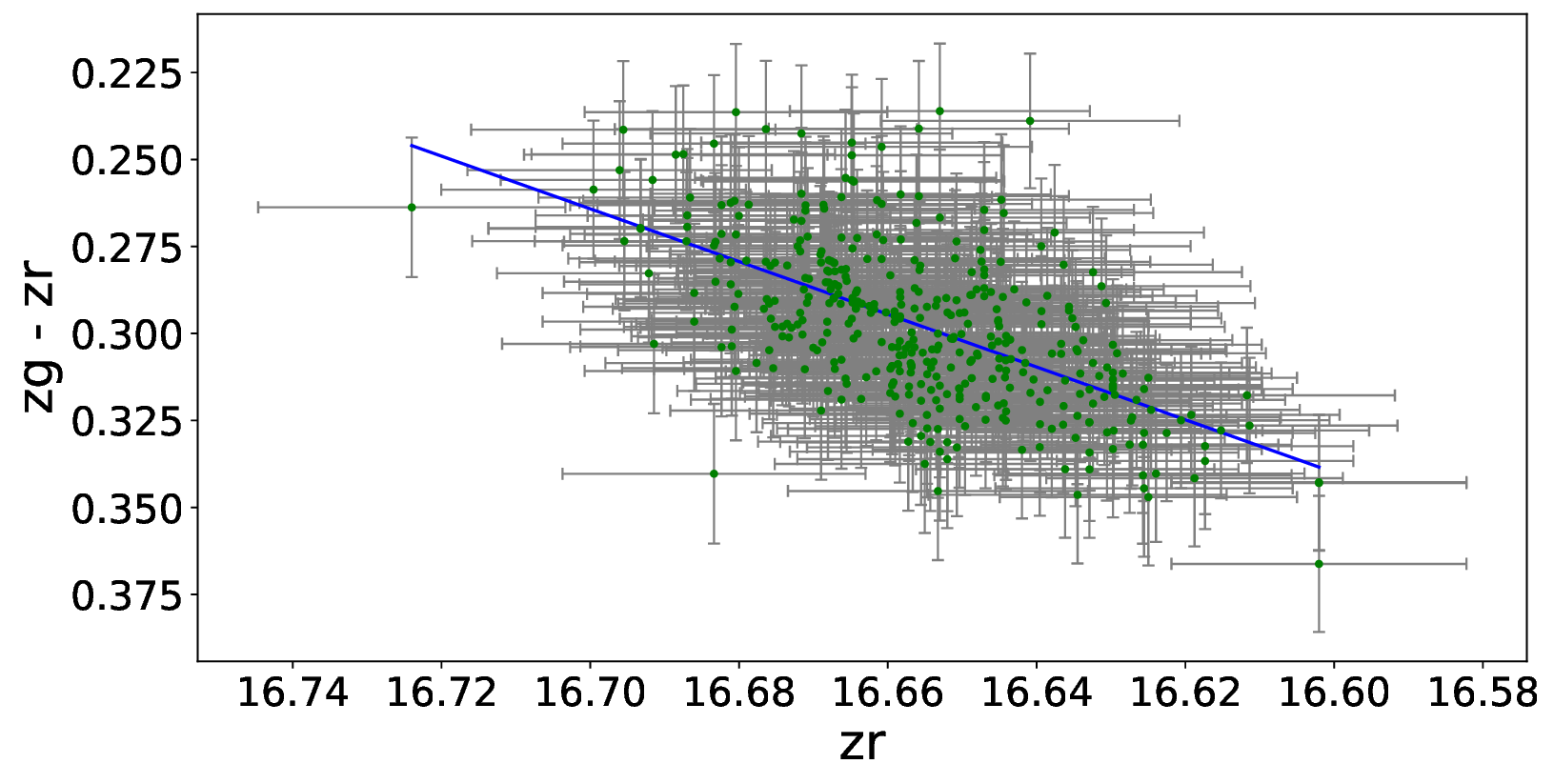}
		\includegraphics[width=0.79\linewidth]{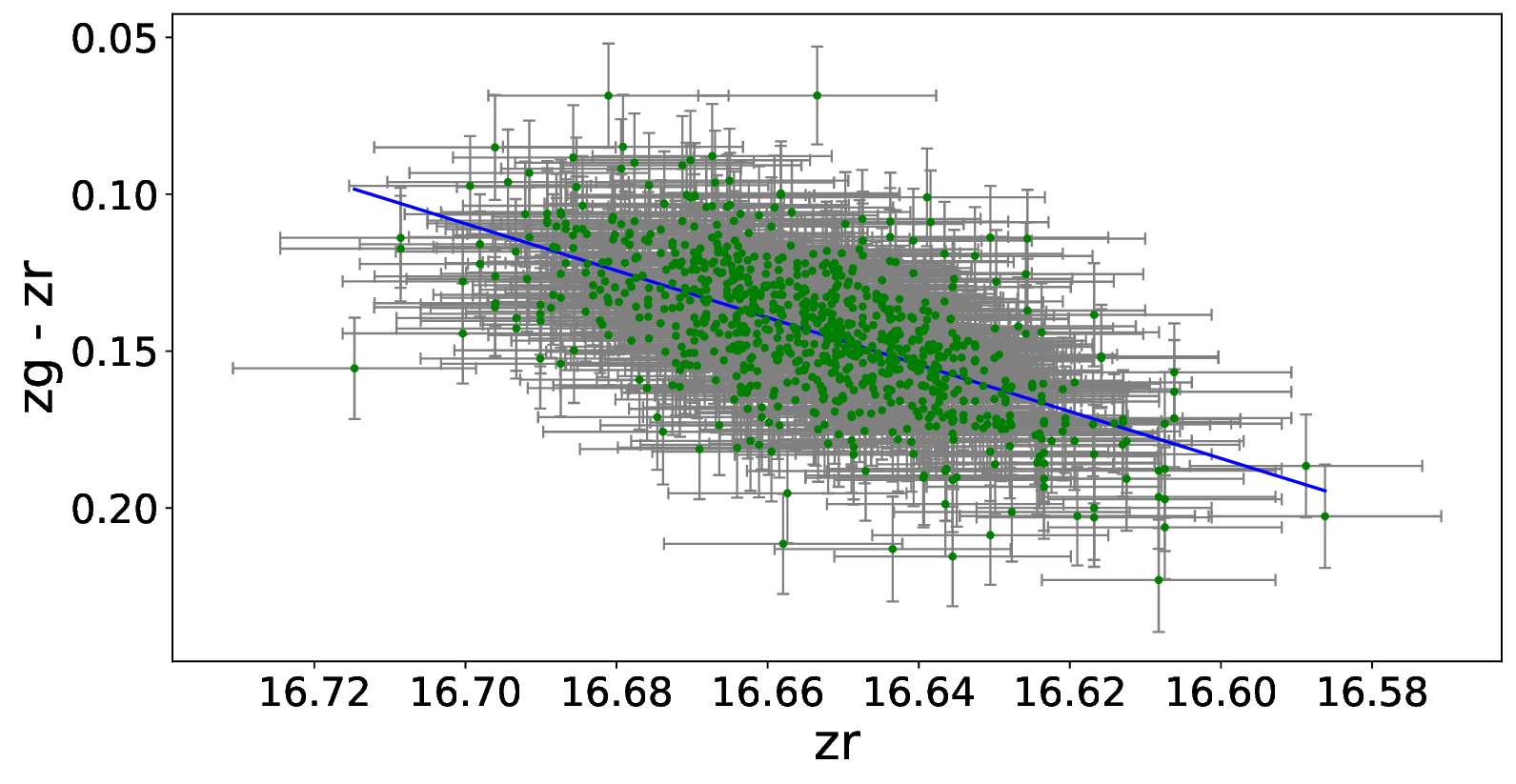}
		\includegraphics[width=0.79\linewidth]{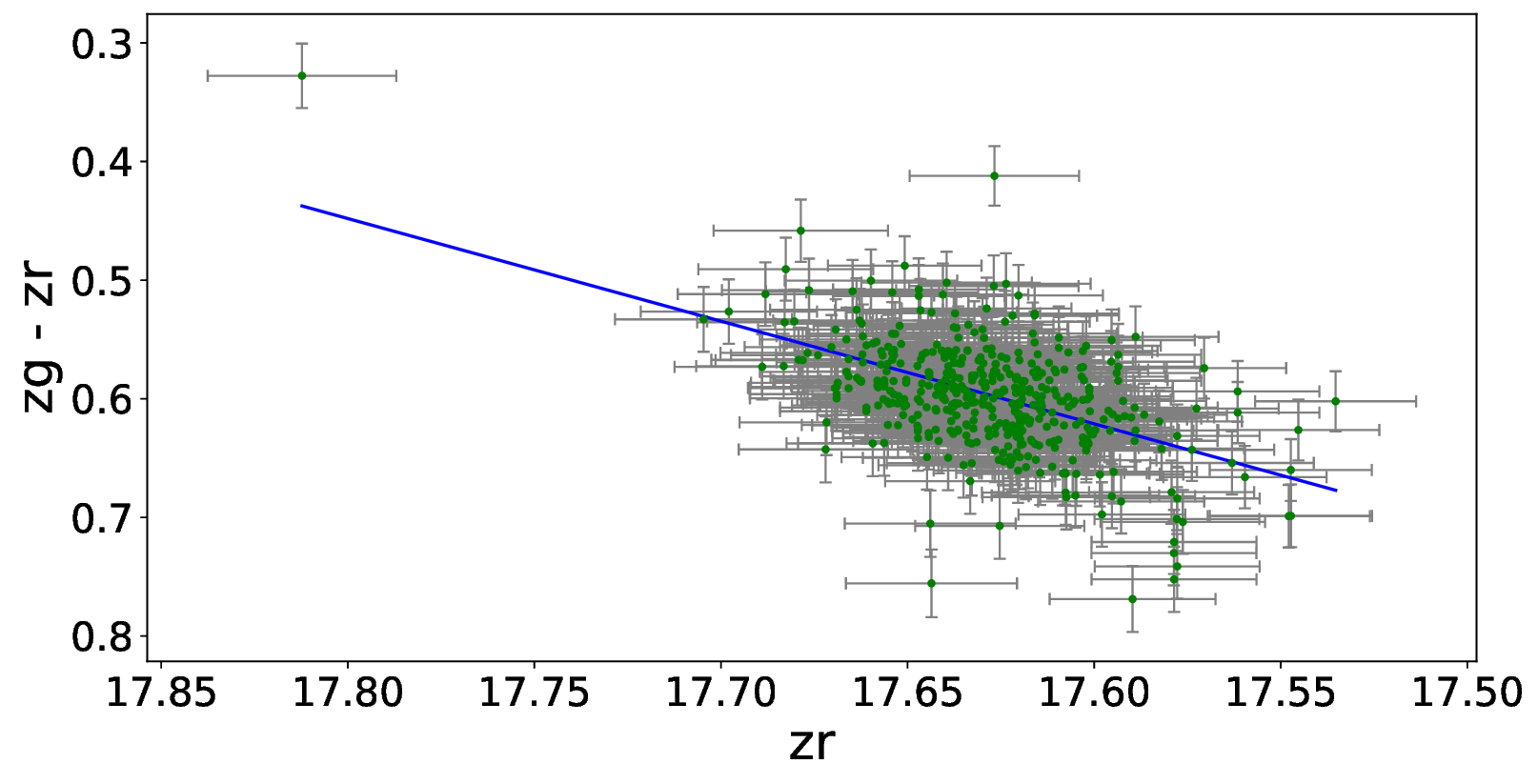}
		\caption{{\it Top} to {\it bottom}: 
			colour-magnitude diagrams of type 1 AGN 
		J1127+2654, J1527+2233, 
		and J1606+2903. Their $k$ values were respectively 
		$-0.758\pm$0.001, $-0.749\pm$0.001, and $-0.865\pm$0.001, and
	$zr$-band variations were less than 0.2\,mag (excluding one
		outlier in J1606+2903).}
		\label{figa:cm}
	\end{figure}

	\section{SPECTRUM FITTING with PYQSOFIT}
	\label{sec:fit}
	\begin{figure*}
		\centering
		\includegraphics[width=0.86\linewidth]{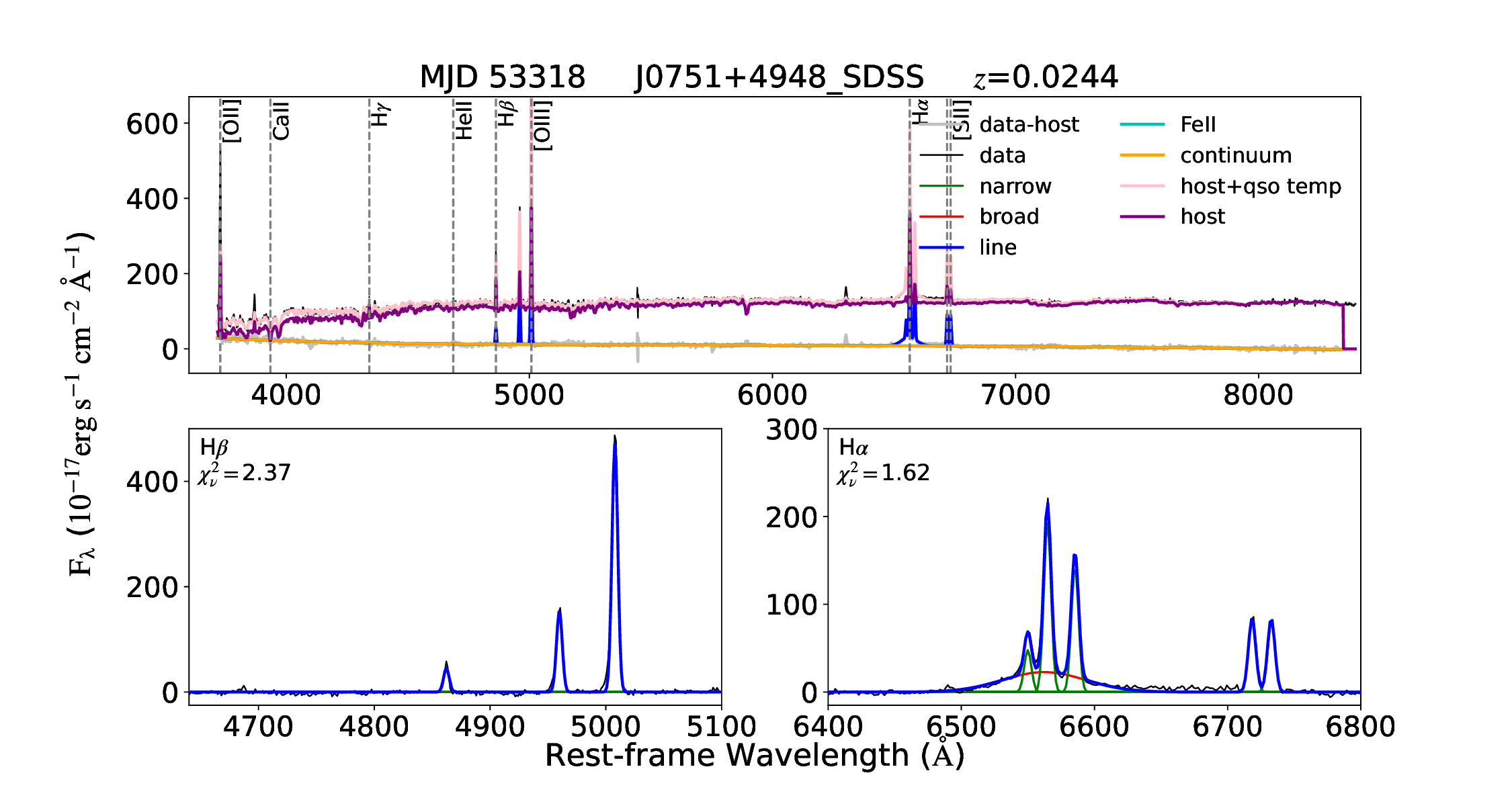}
		\includegraphics[width=0.86\linewidth]{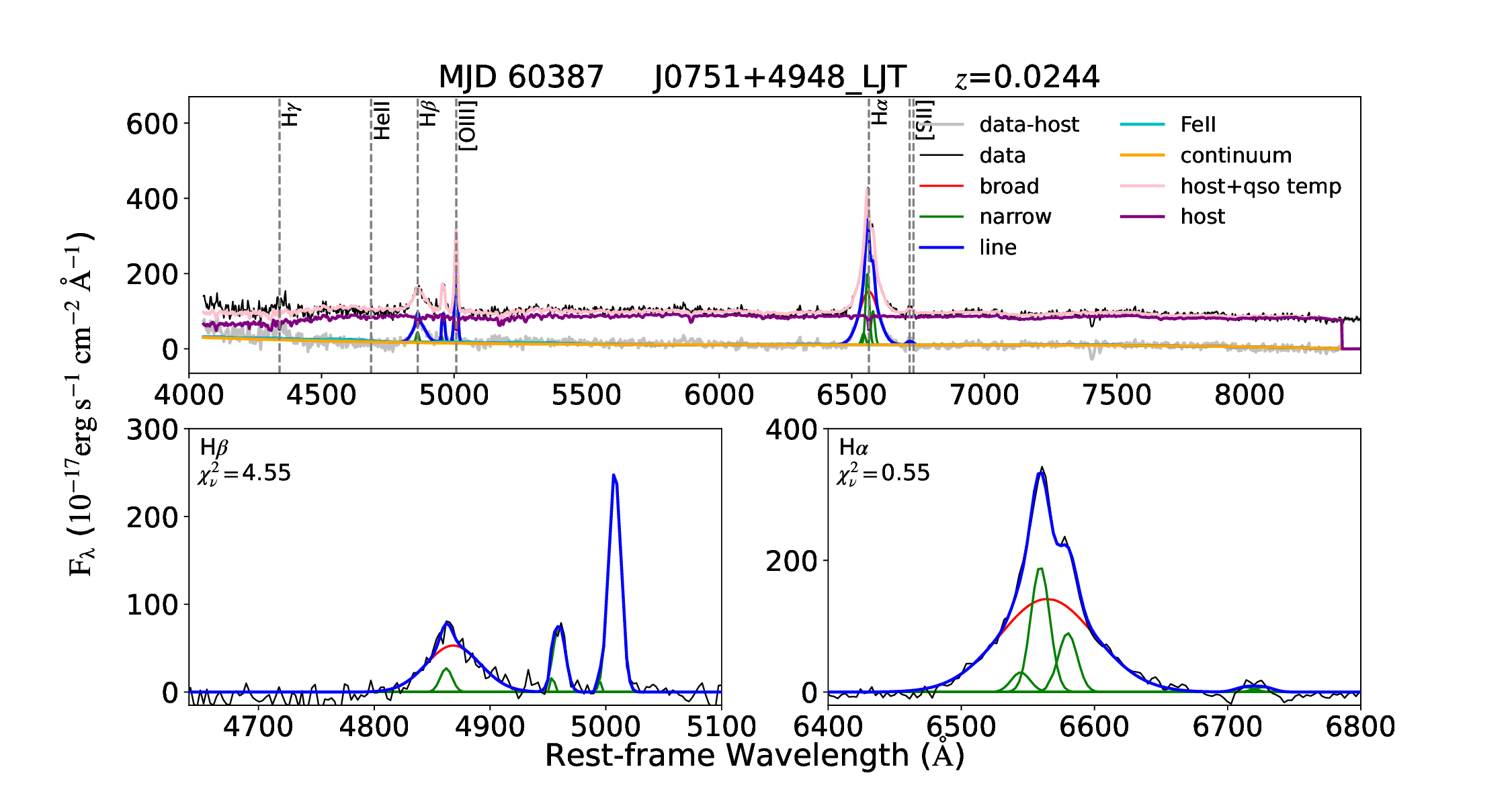}
		\caption{J0751+4948}
		\label{fig:sf1}
	\end{figure*}
	\begin{figure*}
		\includegraphics[width=0.86\linewidth]{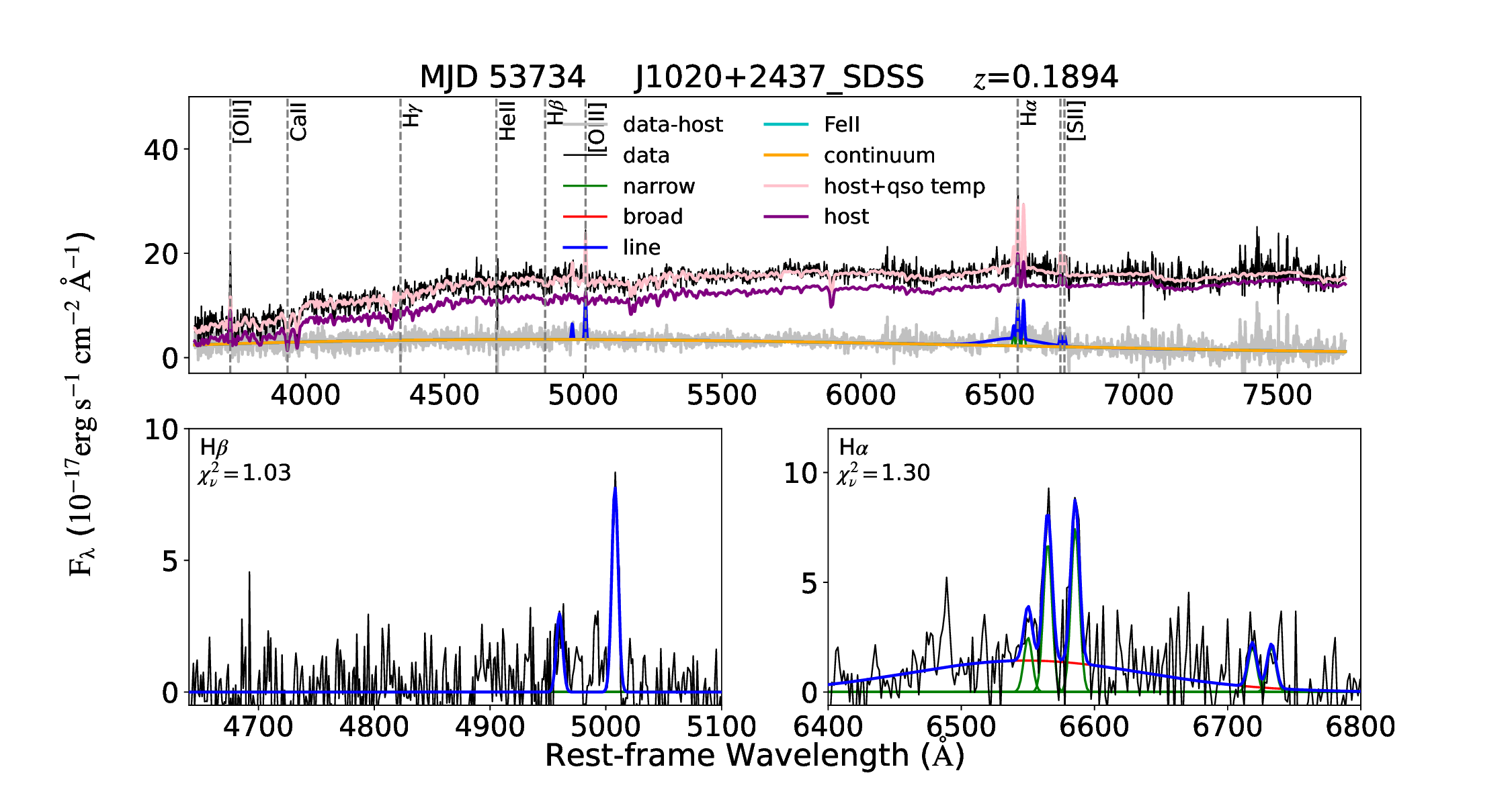}
		\includegraphics[width=0.86\linewidth]{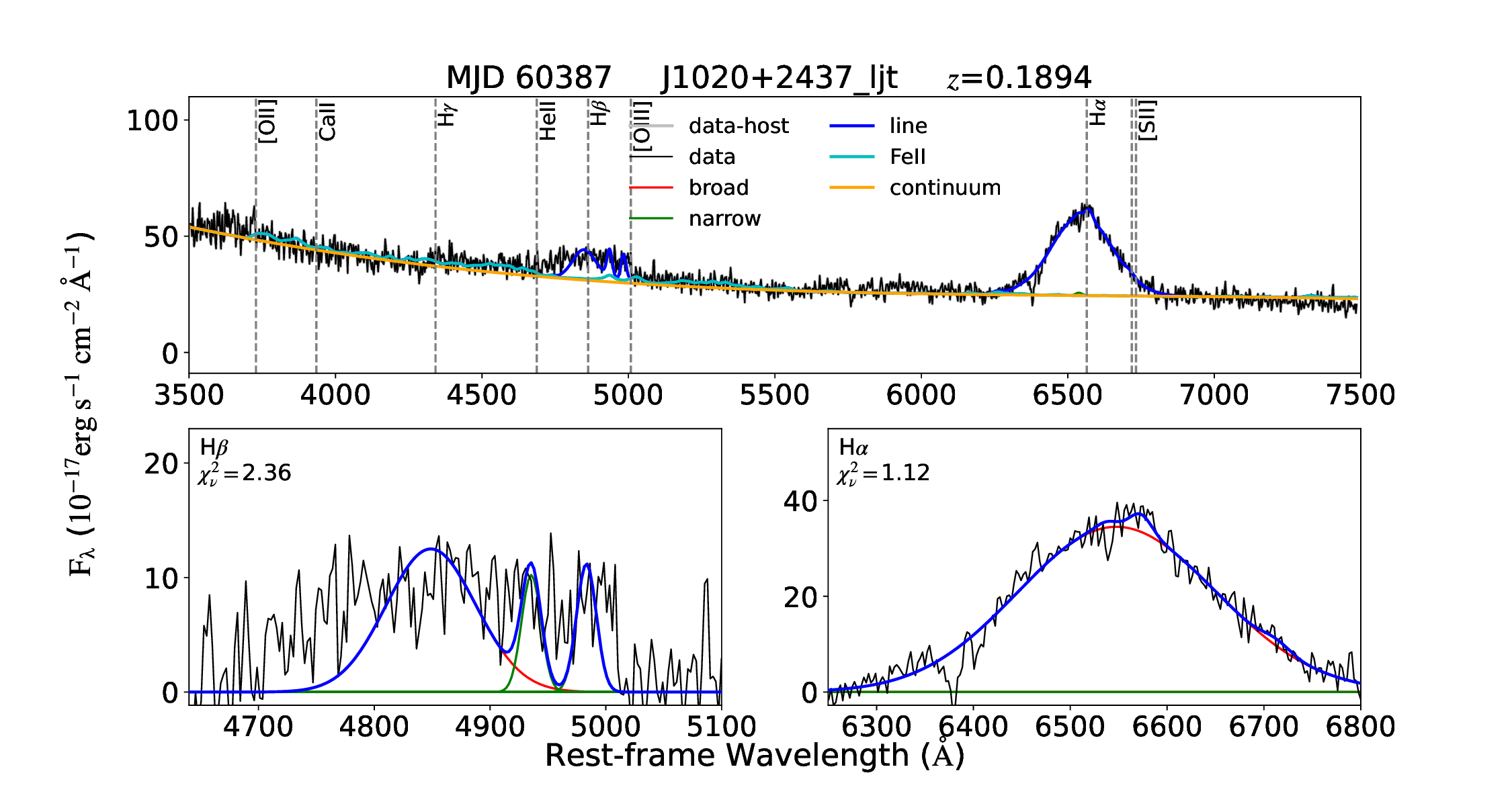}
		\caption{J1020+2437}
		\label{fig:sf2}
	\end{figure*}
	
	\begin{figure*}
		\includegraphics[width=0.86\linewidth]{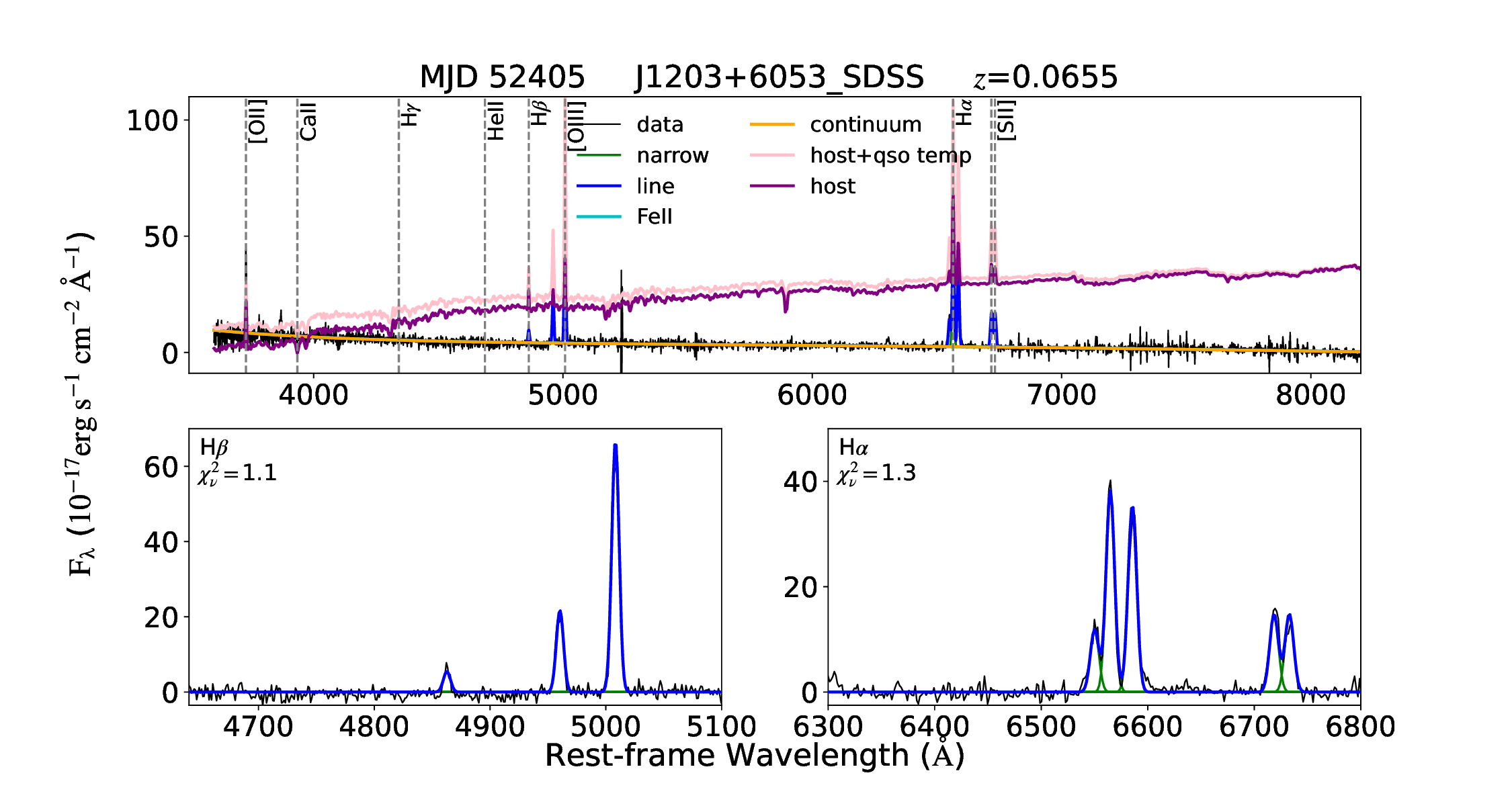}
		\includegraphics[width=0.86\linewidth]{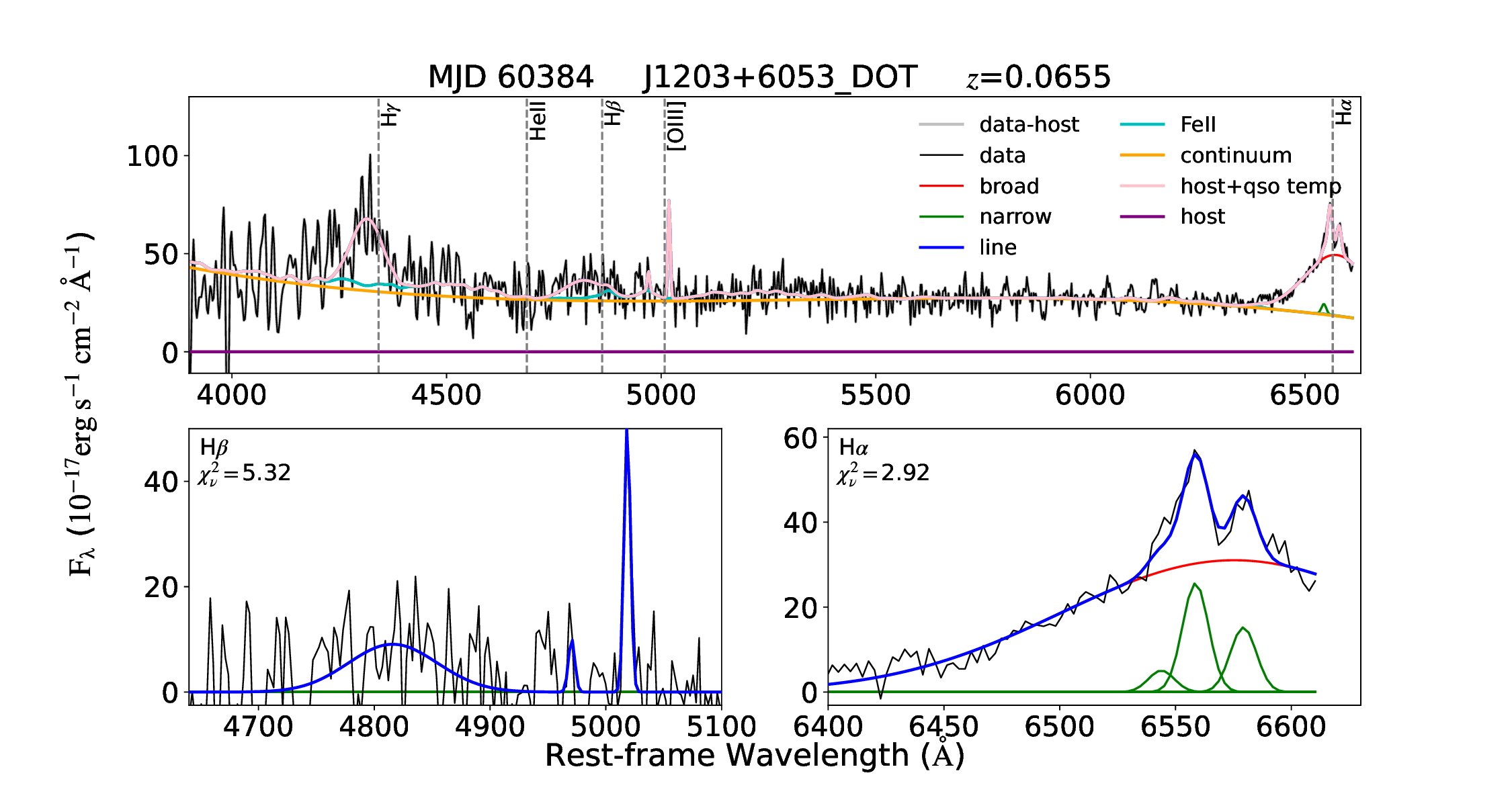}
		\caption{J1203+6053}
		\label{fig:sf3}
	\end{figure*}
	
	\begin{figure*}
		\includegraphics[width=0.86\linewidth]{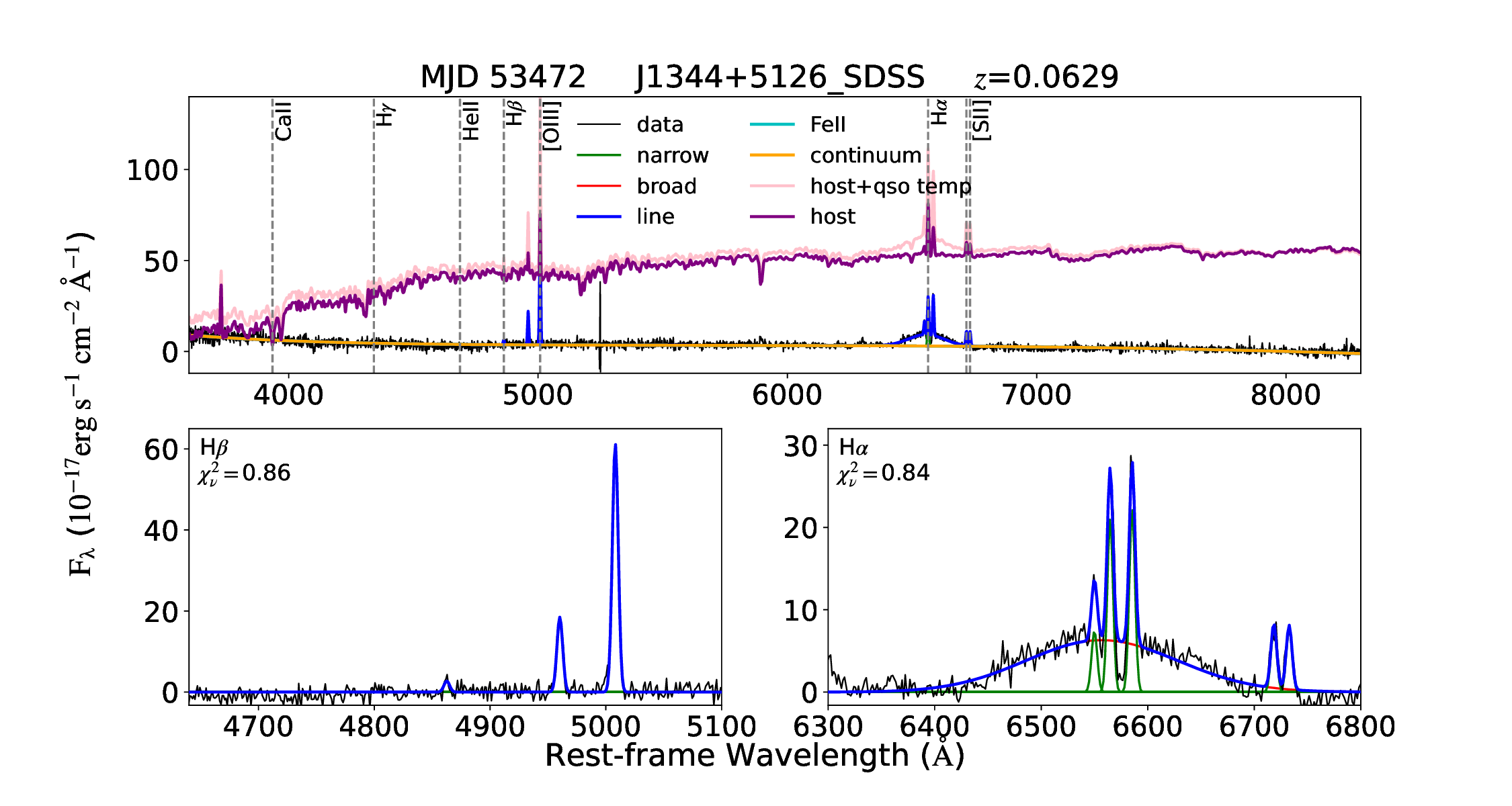}
		\includegraphics[width=0.86\linewidth]{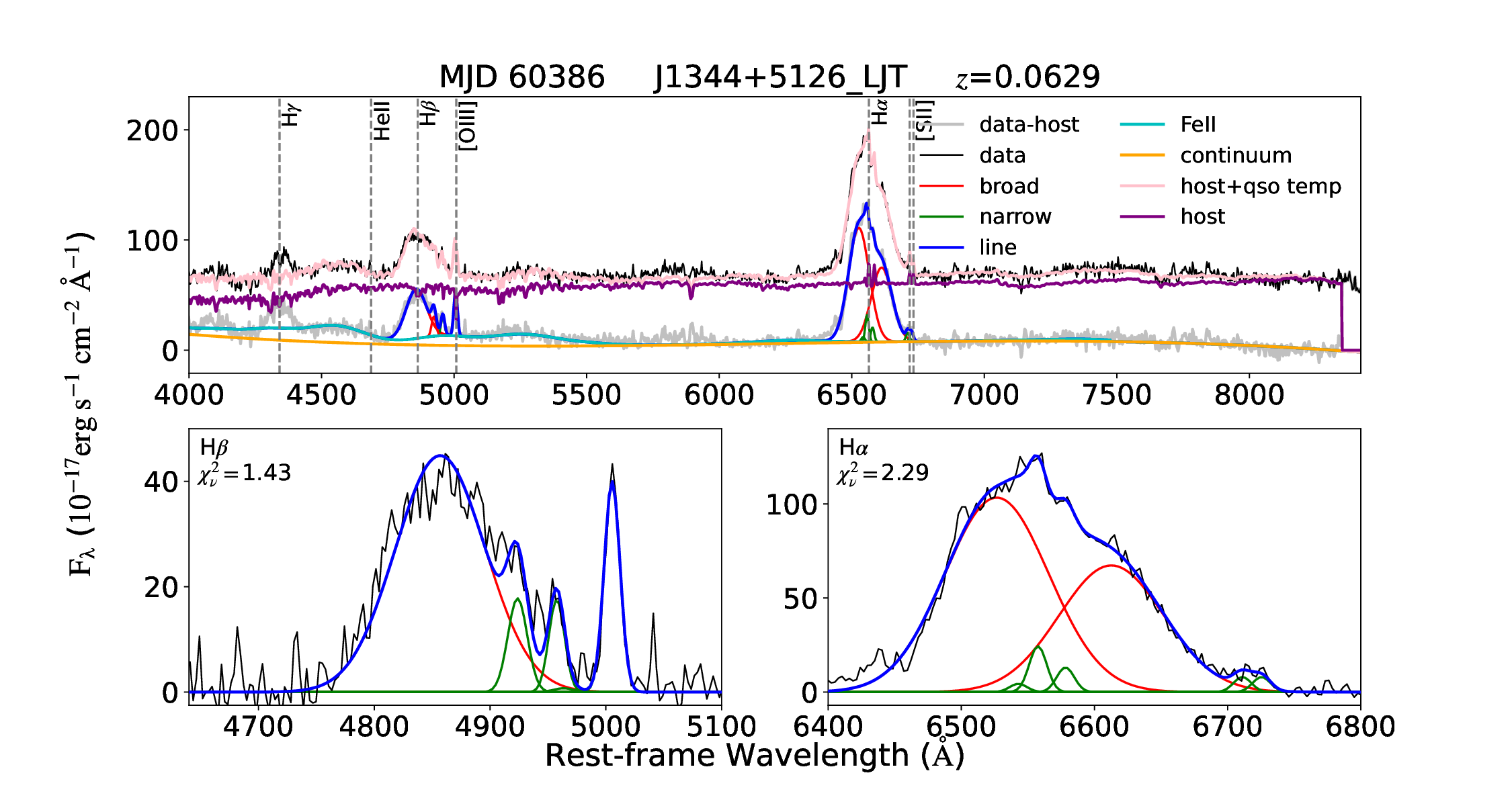}
		\caption{J1344+5126}
		\label{fig:sf4}
	\end{figure*}

\section{Light Curves of Four Type 2 AGNs}
	\begin{figure*}
		 \includegraphics[width=0.51\linewidth]{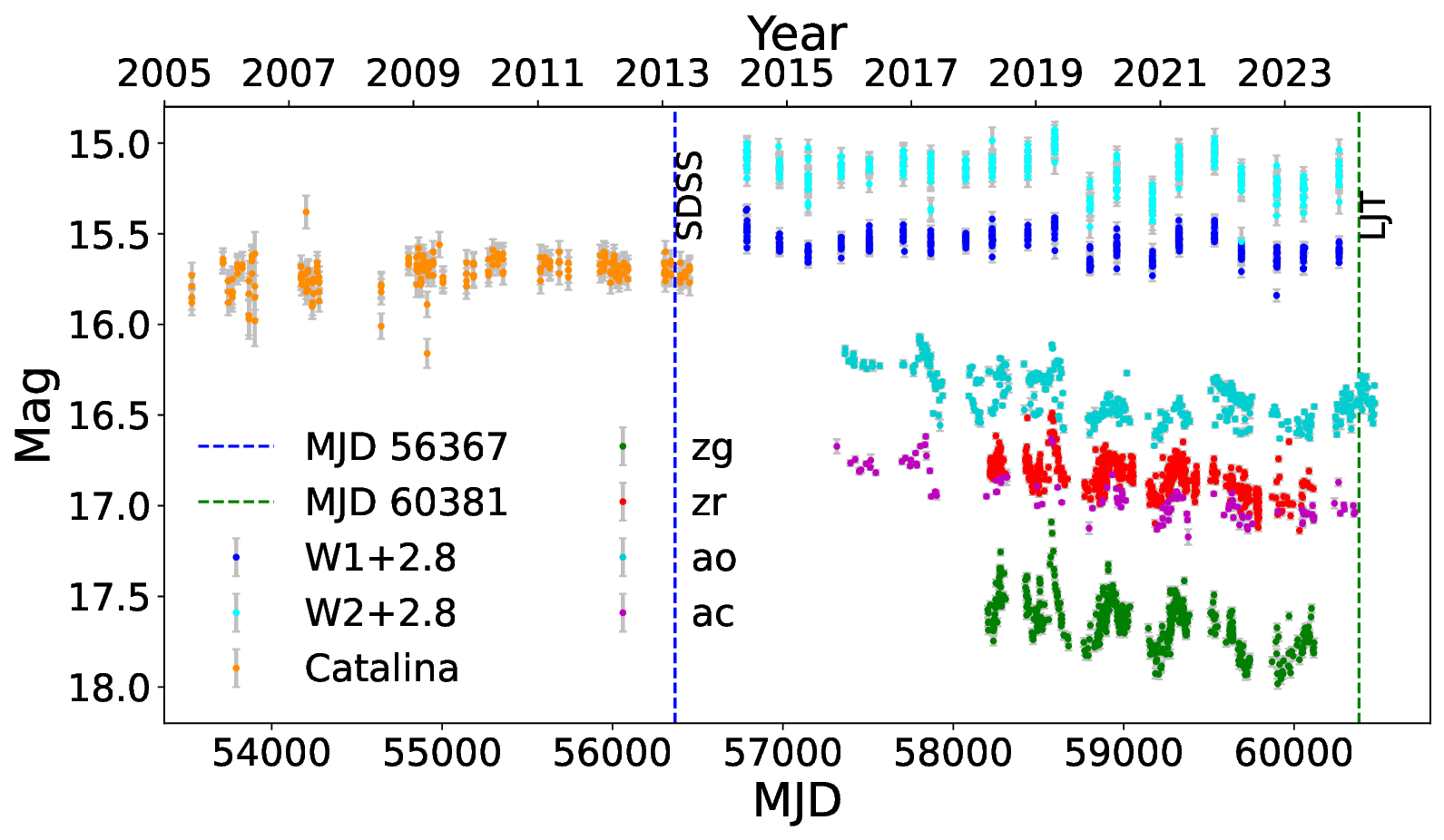}
                \includegraphics[width=0.51\linewidth]{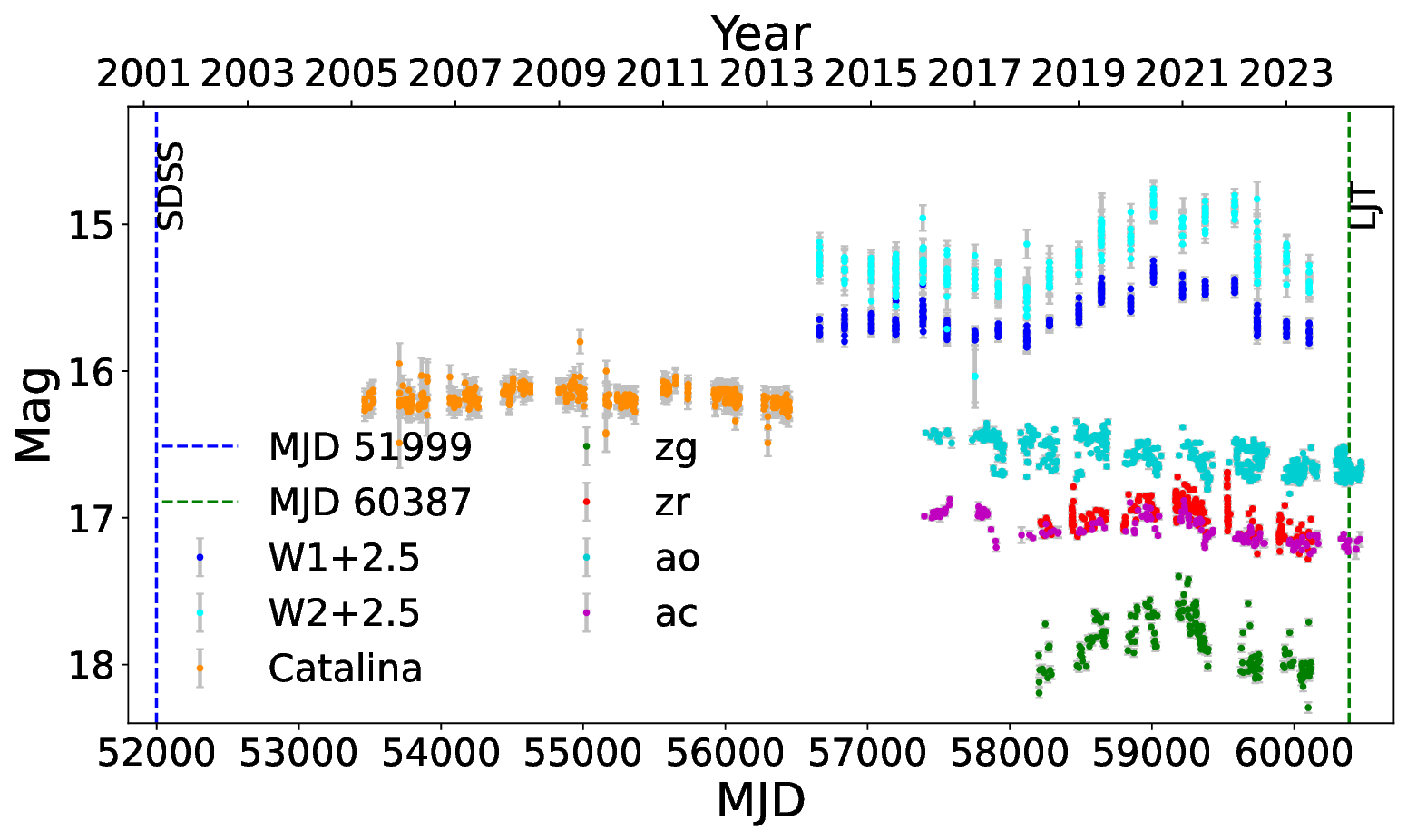}
                \includegraphics[width=0.51\linewidth]{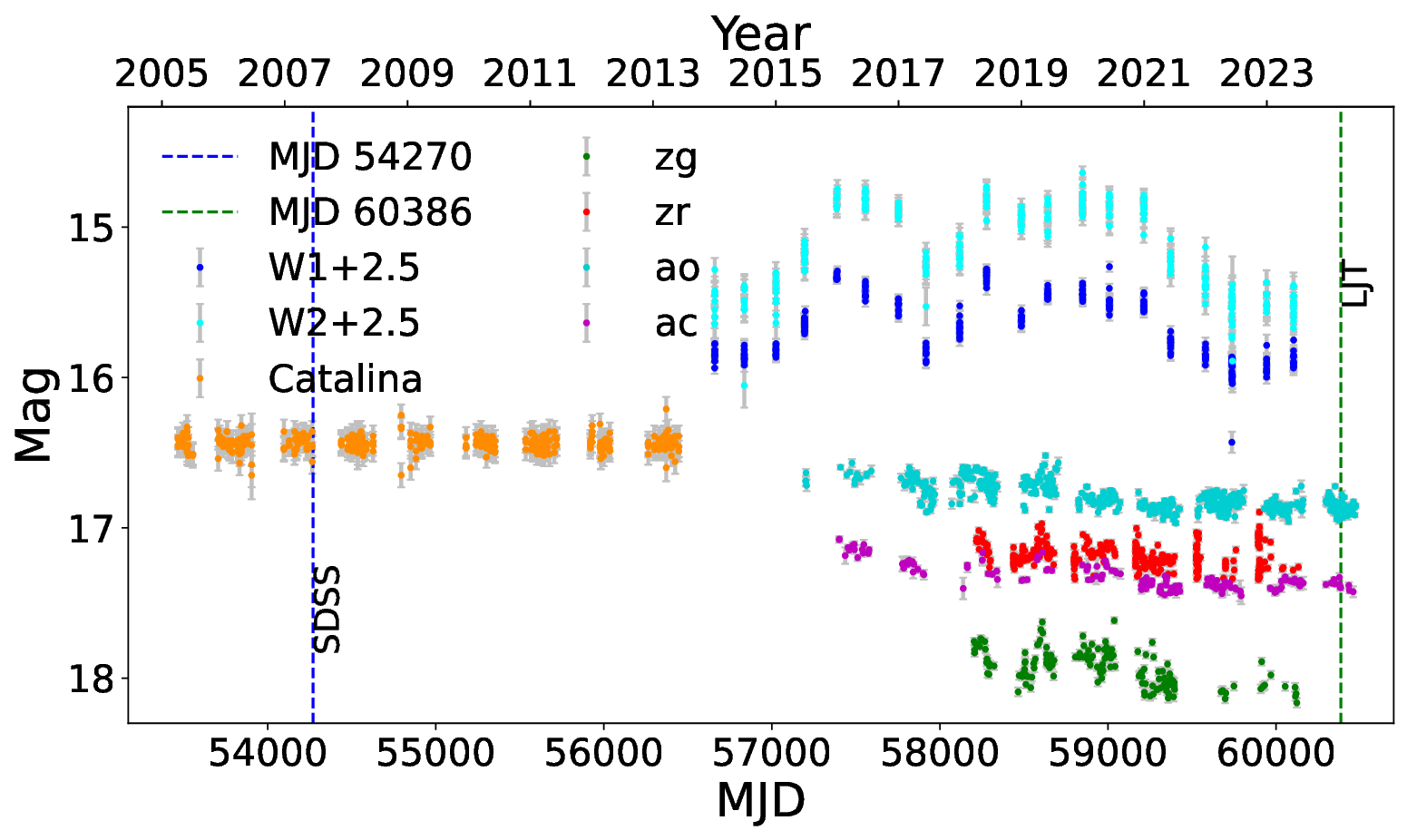}
                \includegraphics[width=0.51\linewidth]{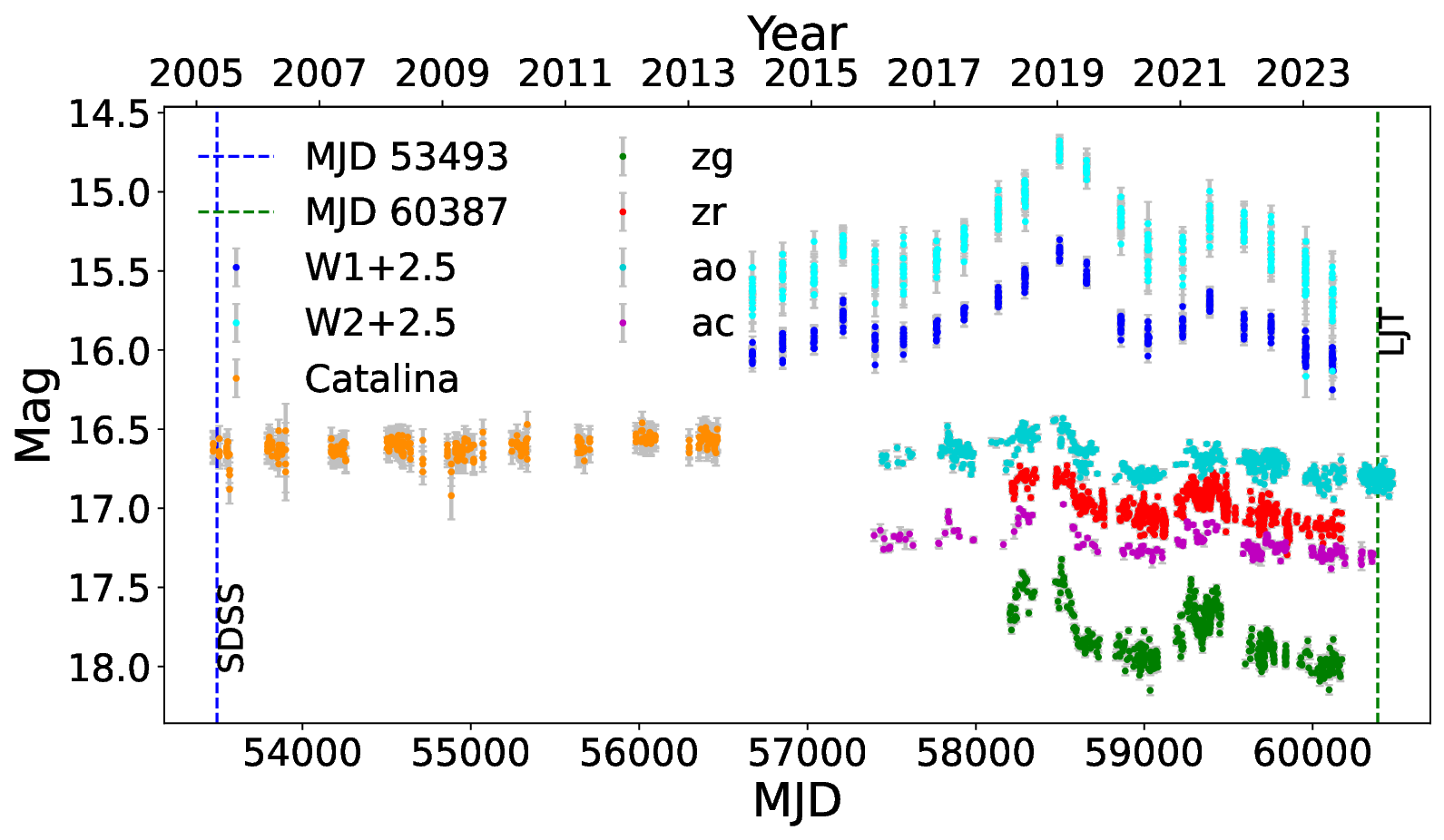}
		\caption{{\it Top} to {\it bottom}: light curves of J1053+4929,
		J1246-0156, J1252+0717, and J1423+2454.}
		\label{figa:four}
	\end{figure*}

	\label{lastpage}
\end{document}